\pdfoutput=1
\documentclass[preprint,12pt]{elsarticle}

\usepackage{amssymb}

\usepackage{amsmath}
\usepackage{float}
\usepackage{booktabs}
\usepackage{array}
\usepackage{tabularx}
\usepackage{multirow}
\usepackage{xcolor}
\usepackage{soul}

\usepackage{algorithm}
\usepackage{algorithmicx}
\usepackage{algpseudocode}
\usepackage{listings}
\usepackage[hidelinks]{hyperref}
\journal{Applied Energy}

\begin{document}

\begin{frontmatter}

\title{Rapid and Robust Parameter Estimation for Electrochemical Battery Models via BOLT: A Batch-Optimized Local-to-Global Technique\tnoteref{t1}}

\tnotetext[t1]{This is the authors' accepted manuscript. \textit{Applied Energy}, 2026, Article 128307, doi: \url{https://doi.org/10.1016/j.apenergy.2026.128307}. The definitive published version is available on ScienceDirect at \url{https://www.sciencedirect.com/science/article/pii/S030626192600961X}. Please cite the published version. \copyright{} 2026 Elsevier Ltd. This manuscript version is made available under the CC BY-NC-ND 4.0 license: \url{https://creativecommons.org/licenses/by-nc-nd/4.0/}.}

\author[inst1,inst2,inst3]{Feng Guo\corref{cor1}}
\author[inst1,inst2]{Luis D. Couto}
\author[inst1,inst2,inst4]{Keivan Haghverdi}
\author[inst1,inst2]{Khiem Trad}
\author[inst1,inst2]{Grietus Mulder}

\affiliation[inst1]{organization={VITO},%Department and Organization
            addressline={Boeretang 200}, 
            city={Mol},
            postcode={2400}, 
            country={Belgium}}

\affiliation[inst2]{organization={EnergyVille},%Department and Organization
            addressline={Thor Park 8310}, 
            city={Genk},
            postcode={3600}, 
            country={Belgium}}
\affiliation[inst3]{%
    organization={Institute for Materials Research (IUMAT), Hasselt University},%
    addressline={Martelarenlaan 42},%
    city={Hasselt},%
    postcode={3500},%
    country={Belgium}%
}
\affiliation[inst4]{organization={Institute of Physical Chemistry, RWTH Aachen University},%Department and Organization
            % addressline={}, 
            city={Aachen},
            postcode={2074}, 
            country={Germany}}

\cortext[cor1]{Corresponding author.}

\begin{abstract}
Accurate and efficient parameter estimation is essential for applying electrochemical battery models in simulation, state estimation, control, and repeated model updating. However, conventional optimization methods, such as particle swarm optimization (PSO) and genetic algorithms (GA), often require many model evaluations and show considerable run-to-run variability, limiting their use in time-sensitive calibration scenarios. This study proposes a Batch-Optimized Local-to-Global Technique (BOLT) for rapid and robust parameter estimation of electrochemical battery models. BOLT combines diversified candidate initialization, batch-parallel trust-region reflective (TRF) local refinement, JIT-accelerated model evaluation, and multi-condition consistency screening within a unified calibration workflow. Comparative experiments based on a grouped single-particle model and measured data from a commercial 18650 NMC lithium-ion cell show that BOLT achieves a favorable trade-off among voltage-response accuracy, computational efficiency, and repeated-run stability. BOLT(32) achieves an average mean absolute error of \(12.4 \pm 0.1\) mV over five operating conditions, requiring only \(20636 \pm 3081\) model calls and \(8.97 \pm 1.20\) s per run. Synthetic-data validation with a known parameter vector in the grouped SPM formulation further shows that BOLT recovers the reference parameter vector under model-consistent conditions and remains robust under 1--3 mV voltage-noise perturbations, with the mean parameter absolute relative error below \(0.6\%\). These results indicate that BOLT provides a practical calibration framework for BMS parameter updating, control-oriented battery digital twins, and second-life battery screening.
\end{abstract}


\end{frontmatter}

\section{Introduction}\label{sec1}

Lithium-ion batteries have become key electrochemical energy storage units in electric vehicles, portable electronics, and stationary energy storage systems \cite{xiong2020research,wang2023system,li2022parameter,cano2018batteries}. As application demands continue to increase, battery models are expected to support not only terminal-voltage prediction, but also high-fidelity simulation, state estimation, fault diagnosis, control, and lifecycle management \cite{haghverdi2025review,qi2024battery,konz2023high,hu2020battery,gao2022development}. In particular, the rapid development of digital twins \cite{dubarry2023enabling,naseri2023digital,song2025microstructural} and world-model-related intelligent frameworks \cite{hafner2025mastering} is further increasing the demand for accurate and computationally efficient physics-based battery models. Among the available modeling approaches, electrochemical models are particularly attractive because they can describe internal reaction and transport mechanisms and therefore provide richer physical interpretability than conventional equivalent circuit models (ECMs) \cite{zhu2026reduced,ali2024comparison,mama2025comprehensive}. However, the practical deployment of electrochemical models remains strongly constrained by parameter estimation, which is often computationally demanding, ill-conditioned, and difficult to perform robustly across repeated runs \cite{miguel2021review,li2024comparative,ding2023accurate,cai2024characterization}.

Electrochemical battery models generally refer to the Pseudo-Two-Dimensional (P2D) model and its simplified variants \cite{fuller1994simulation}. Common reduced-order forms derived from the P2D framework include the Single Particle Model with electrolyte (SPMe) \cite{marquis2019asymptotic} and the Single Particle Model (SPM) \cite{haran1998determination}. Compared with ECMs, these models incorporate internal physicochemical states such as lithium concentration distributions and electrode reaction kinetics, which makes them more suitable for advanced battery management tasks requiring physical consistency and interpretability \cite{guo2024systematic}. However, this advantage comes at the price of a substantially larger and more coupled parameter set. Some parameters require destructive characterization procedures, such as cell disassembly or half-cell measurements, whereas others cannot be directly measured at all \cite{ecker2015parameterization,lu2017state,wang2021decoupling}. As a result, parameter estimation based on measurable current--voltage--temperature data has become a central problem in electrochemical model deployment. Yet because electrochemical models are highly over-parameterized, the identification problem often suffers from non-uniqueness, meaning that distinct parameter sets may produce similar output behavior \cite{rojas2024critical}. Together with the high computational burden of repeated model simulation, this makes fast, stable, and practically deployable parameter estimation a persistent challenge.

A first major line of research relies on local gradient-based or least-squares-type methods for electrochemical model parameter estimation \cite{deng2017implementation}. Such methods are attractive because they are generally efficient and can converge rapidly once initialized near a good solution. However, their performance is strongly dependent on the initial guess, and in the nonconvex parameter landscape of electrochemical models they may converge to poor local minima. Therefore, although local solvers are computationally appealing, their repeated-run robustness is often insufficient for practical calibration tasks where reliable parameter estimation must be obtained repeatedly under limited time budgets.

A second major line of research uses population-based metaheuristic optimization methods, such as particle swarm optimization (PSO), genetic algorithms (GA), and their variants \cite{rahman2016electrochemical,xiong2018electrochemical}. Our previous work compared several classical methods and showed that PSO provided relatively favorable performance in terms of stability and accuracy among the tested baselines \cite{guo2024efficiency}. Other studies have introduced increasingly sophisticated metaheuristics. For example, Tian et al. used a Two-Population Grey Wolf Optimization method for P2D parameter identification, but the reported optimization required about 12 hours \cite{tian2025physics}. Huang et al. proposed a multi-step Cuckoo Particle Swarm Optimization method for SPMe parameter estimation, but the total computational cost was not clearly reported \cite{huang2024novel}. These studies demonstrate that stronger global search can improve calibration quality, but they also reveal an important limitation: population-based methods typically require a large number of repeated battery model evaluations, which leads to high computational cost and limits their practicality in repeated or time-sensitive calibration scenarios.

A third line of work seeks to accelerate parameter estimation through surrogate modeling or machine learning. Kim et al. introduced a deep Bayesian neural network for electrochemical model parameter estimation \cite{kim2019data}. Wang et al. proposed a classification-model-assisted Bayesian optimization strategy that improved estimation accuracy relative to PSO and GA while reducing computation time compared with GA \cite{wang2024fast}. Zhang et al. developed a multi-step meta-modeling genetic algorithm in which a neural network surrogate is trained to predict voltage responses and then used to accelerate optimization \cite{zhang2025multi}. More broadly, Li et al. compared 78 metaheuristic algorithms and proposed a modified Teaching-Learning-Based Optimization approach with improved stability and runtime \cite{li2024comparative}. These studies clearly indicate that computational acceleration is possible. However, many surrogate- or learning-assisted approaches introduce additional workflow complexity, such as surrogate training, tuning, and reliability management. As a result, although they may reduce part of the computational burden, they do not always provide a simple and directly deployable repeated-calibration framework for electrochemical battery models.

Taken together, existing studies have improved electrochemical model parameter estimation mainly by strengthening global search, accelerating model evaluation, or introducing surrogate assistance. However, a practical gap remains. Local gradient-based methods are computationally efficient but sensitive to initialization; population-based metaheuristics provide broader search capability but require many repeated model evaluations; and surrogate- or learning-assisted approaches can reduce part of the computational burden but often introduce additional training, tuning, and reliability-management complexity. Therefore, the literature still lacks a simple and practical calibration framework that can simultaneously provide fast computation, repeated-run robustness, and cross-condition consistency for deployment-oriented electrochemical battery model calibration.

To address this gap, this study proposes a Batch-Optimized Local-to-Global Technique (BOLT) for rapid and robust parameter estimation of electrochemical battery models. The core idea is to reallocate the computational budget from expensive population evolution to diversified local refinement and multi-condition candidate screening. Specifically, BOLT first generates multiple physically constrained candidate parameter vectors, refines them independently using TRF-based local optimization, and then selects the final parameter set according to its averaged voltage error over multiple operating conditions. This design aims to combine the computational efficiency of local solvers with the robustness of multi-candidate exploration and cross-condition screening.

The main contributions of this work are fourfold. First, BOLT develops a batch-optimized local-to-global calibration framework that combines diversified candidate initialization with TRF-based local refinement for electrochemical battery model parameter estimation. Second, a batch-parallel execution strategy is introduced to improve computational efficiency by distributing independent local refinements across available vCPUs. Third, a multi-condition consistency screening mechanism is incorporated to select the candidate parameter set with the lowest averaged condition-wise voltage error, reducing dependence on a single fitting condition. Fourth, the evaluation combines repeated-run analysis, candidate-number sensitivity analysis, measured-data comparison, and synthetic known-parameter tests with voltage-noise perturbations, thereby assessing both voltage-response accuracy and parameter recovery capability.

\section{Electrochemical model}
This section introduces the SPM, presents its discretized grouped formulation, and defines the parameter set considered in the estimation problem.

\subsection{SPM Fundamentals}

\begin{figure}[H]
\centering
\makebox[\textwidth][c]{
    \includegraphics[width=0.8\textwidth]{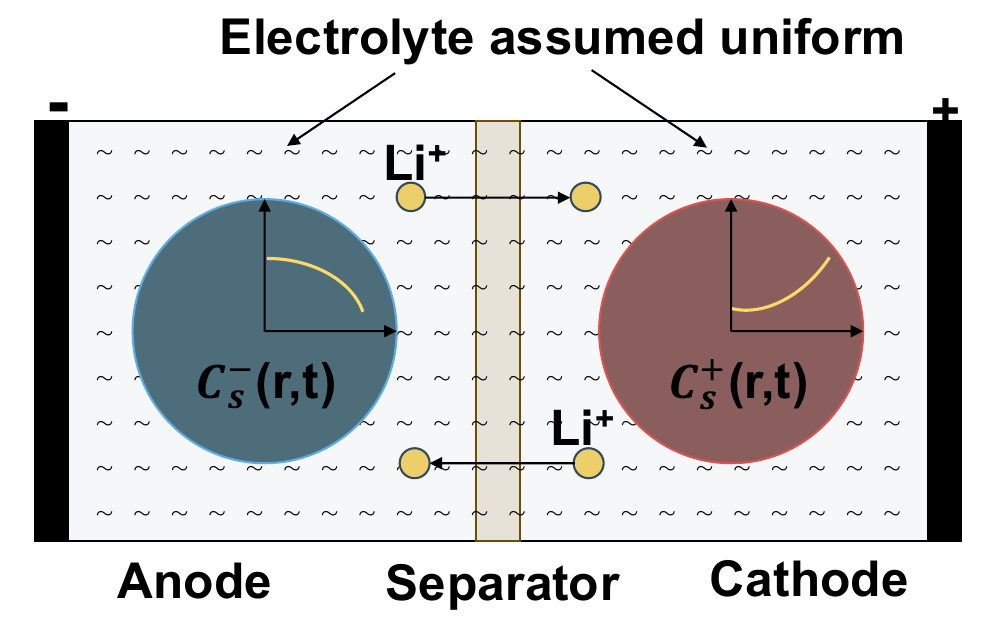}
}
\caption{Schematic illustration of the SPM considered in this study.}
\label{fig:1}
\end{figure}

The electrochemical model employed in this study is the SPM (as shown in Figure~\ref{fig:1}), which is frequently adopted in research due to its relatively simple structure and ease of implementation. The proposed framework can also be extended to more complex electrochemical models. The solid-phase lithium concentration inside a representative spherical particle evolves according to
\begin{equation}
\frac{\partial c_{s,i}}{\partial t}(r,t)=\frac{D_{s,i}}{r^{2}}\,
\frac{\partial}{\partial r}\!\left(r^{2}\frac{\partial c_{s,i}}{\partial r}(r,t)\right),
\label{eq:1}
\end{equation}
where \(i\in\{p,n\}\) denotes the positive (\(p\)) and negative (\(n\)) electrodes, \(c_{s,i}\) is the solid-phase concentration, and \(D_{s,i}\) the diffusion coefficient.
A spatially uniform initial concentration is assumed, i.e.\ \(c_{s,i}(r,0)=c_i(0)\).

The boundary conditions accompanying~\eqref{eq:1} are
\begin{align}
\left.\frac{\partial c_{s,i}}{\partial r}\right|_{r=0}&=0, \label{eq:2}\\[4pt]
\left.\frac{\partial c_{s,i}}{\partial r}\right|_{r=R_{s,i}}&=-\frac{j_i(t)}{D_{s,i}}, \label{eq:3}
\end{align}
enforcing symmetry at the particle centre and prescribing the intercalation flux \(j_i(t)\) at the particle surface, where \(R_{s,i}\) denotes the particle radius.

The cell voltage predicted by the SPM reads
\begin{equation}
\begin{aligned}
V_{\text{SPM}}(t)=
  OCP_p\!\bigl(\tilde{c}_{ss,p}(t)\bigr)-
  OCP_n\!\bigl(\tilde{c}_{ss,n}(t)\bigr) \\
  +\eta_p\!\bigl(\tilde{c}_{ss,p}(t),I(t)\bigr)-
   \eta_n\!\bigl(\tilde{c}_{ss,n}(t),I(t)\bigr)-R_0\,I(t),
\end{aligned}
\label{eq:4}
\end{equation}
where \(OCP_i\) is the open-circuit potential of electrode \(i\) (expressions are given in Appendix~A), \(\eta_i\) the surface over-potential, \(I(t)\) the applied current and \(R_0\) a lumped ohmic resistance.
The normalised surface stoichiometry is defined as
\begin{equation}
\tilde{c}_{ss,i}(t)=\frac{c_{ss,i}(t)}{c_{\text{max},i}},
\label{eq:5}
\end{equation}
with \(c_{ss,i}(t)=c_{s,i}(R_{s,i},t)\) and \(c_{\text{max},i}\) the maximum solid-phase concentration.

The over-potential for electrode \(i\) is given by
\begin{equation}
\eta_i\!\bigl(\tilde{c}_{ss,i},I\bigr)=
\frac{2RT}{F}\,
\sinh^{-1}\!\left(
\frac{1_{\mp}\,I}{2\,a_i\,L_i\,j_{0,i}(\tilde{c}_{ss,i})}
\right),
\label{eq:6}
\end{equation}
where \(R\) is the universal gas constant, \(T\) the absolute temperature, \(F\) Faraday’s constant, and the sign factor \(1_{\mp}\) equals \(-1\) for the positive and \(+1\) for the negative electrode.
The exchange current density is
\begin{equation}
j_{0,i}(\tilde{c}_{ss,i})=
r_{\text{eef},i}\,c_{\text{max},i}\,
\sqrt{c_e\,\tilde{c}_{ss,i}\,\bigl(1-\tilde{c}_{ss,i}\bigr)},
\label{eq:7}
\end{equation}
with \(r_{\text{eef},i}\) the reaction-rate constant and \(c_e\) the electrolyte lithium-ion concentration.

Assuming a uniform reaction rate across the electrode thickness, the intercalation flux becomes
\begin{equation}
j_i(t)=1_{\mp}\,\frac{I(t)}{F\,a_i\,A_i\,L_i},
\label{eq:8}
\end{equation}
where \(A_i\) and \(L_i\) are the electrode surface area and thickness, respectively.
The specific surface area is evaluated from the solid-phase volume fraction \(\varepsilon_{s,i}\) as
\begin{equation}
a_i=\frac{3\,\varepsilon_{s,i}}{R_{s,i}}.
\label{eq:9}
\end{equation}

Because \(R\) and \(F\) are physical constants and \(c_e\) is normally fixed (e.g.\ 1000 mol m\(^{-3}\)), they do not require estimation.
Consequently, the parameter vector to be identified contains 17 elements:
\begin{equation}
\begin{aligned}
P_{\text{all}}=\bigl[
&D_{s,n},\,D_{s,p},\,R_{s,n},\,R_{s,p},\,
  r_{\text{eef},n},\,r_{\text{eef},p},\\
&A_n,\,A_p,\,L_n,\,L_p,\,
  c_{\text{max},n},\,c_{\text{max},p},\\
&\varepsilon_{s,n},\,\varepsilon_{s,p},\,
  c_n(0),\,c_p(0),\,R_0
\bigr].
\end{aligned}
\label{eq:10}
\end{equation}

Equation~(\ref{eq:10}) gives the full physical parameterization of the SPM. In the following subsection, this formulation is rewritten using a parabolic-approximation-based grouped representation, which leads to a reduced parameter set for the actual estimation problem.

\subsection{Model Discretization }

Due to the presence of partial differential equations (PDEs) in electrochemical battery models (Eq.~\ref{eq:1}), the numerical method used to solve these PDEs can significantly affect the overall speed of parameter estimation. In this study, we adopt the parabolic approximation method to solve the electrochemical model. This approach offers a favorable balance between computational efficiency and accuracy, making it particularly suitable for scenarios requiring rapid model evaluations
\cite{guo2025comparative}. The grouped SPM formulation used in this study is based on our previous work~\cite{guo2025control, guo2026cpg}. For detailed model descriptions, readers are referred to our earlier publication~\cite{guo2025control}. The state-space representation and the parameters to be estimated are as follows:

\begin{align}
    \label{eq:11}
    \dot{\chi}(t) & = \tilde{A} \chi(t) + \tilde{B} u(t), \quad \chi(0) = \text{SOC}_{i}(0), \\
    \label{eq:12}
    \psi(t) & = \tilde{C} \chi(t) + \tilde{D} u(t),
\end{align}
where the normalized state vector is defined as \( \chi = \begin{bmatrix} \chi_p^\top & \chi_n^\top \end{bmatrix}^\top \), with each subvector given by \( \chi_i = \begin{bmatrix} \tilde{q}_{1,i} & \tilde{q}_{2,i} \end{bmatrix}^\top \), \( i \in \{p,n\} \) corresponds to the positive and negative electrodes, respectively. The system input is the applied current \( u(t) = I(t) \), and the output vector \( \psi = \begin{bmatrix} \psi_p^\top & \psi_n^\top \end{bmatrix}^\top \), where each \( \psi_i = \begin{bmatrix} \tilde{\overline{c}}_{s,i} & \tilde{c}_{ss,i} \end{bmatrix}^\top \). \(\tilde{c}_{ss,i}\) is the normalized surface lithium concentration, and \(\tilde{\overline{c}}_{s,i}\) is the normalized average lithium concentration.

The state matrices are structured as:
\begin{equation}
\tilde{A} = \text{diag}(\tilde{A}_p, \tilde{A}_n), \quad
\tilde{B} = \begin{bmatrix} \tilde{B}_p^\top & \tilde{B}_n^\top \end{bmatrix}^\top,
\label{eq:13}
\end{equation}
with each electrode-specific block defined by:
\begin{equation}
\tilde{A}_i = \begin{bmatrix}
0 & 0 \\
\displaystyle \frac{30}{\alpha_i} & \displaystyle -\frac{30}{\alpha_i}
\end{bmatrix}, \quad
\tilde{B}_i = \begin{bmatrix}
\displaystyle \frac{1}{b_i} \\
\displaystyle \frac{19}{7 b_i}
\end{bmatrix},
\label{eq:14}
\end{equation}

The corresponding output matrices are given by:
\begin{equation}
\tilde{C} = \text{diag}(\tilde{C}_p, \tilde{C}_n), \quad
\tilde{D} = \begin{bmatrix} \tilde{D}_p^\top & \tilde{D}_n^\top \end{bmatrix}^\top,
\label{eq:15}
\end{equation}
where:
\begin{equation}
\tilde{C}_i = \begin{bmatrix}
1 & 0 \\
0 & 1
\end{bmatrix}, \quad
\tilde{D}_i = \begin{bmatrix}
0 \\
\displaystyle \frac{\alpha_i}{105 b_i}
\end{bmatrix}.
\label{eq:16}
\end{equation}

The grouped parameters are defined as \( \alpha_i = {R_{s,i}^2}/{D_{s,i}} \) and \( b_i = F A_i L_i \varepsilon_{s,i} c_{{\rm max},i} \), where \( b_i \) represents the capacity-related grouped parameter of electrode \(i\).

Using the grouped formulation, the overpotential term in Eq.~(\ref{eq:6}) can be rewritten as:

\begin{equation}
\eta_{i}(\tilde{c}_{ss,i}(t), I(t)) = \frac{2RT}{F} \sinh^{-1} \left( \frac{1_\mp I(t)}{6 b_i d_i \sqrt{\tilde{c}_{ss,i}(t)(1 - \tilde{c}_{ss,i}(t))}} \right),
\label{eq:17}
\end{equation}
where:
\begin{equation}
d_i = \frac{r_{\text{eef},i}  \sqrt{c_e}}{F R_{s,i}}.
 \label{eq:44}
\end{equation}
Consequently, the parameters requiring estimation are listed in Table~\ref{tab:1}.

Accordingly, the actual estimation problem considered in this work is formulated in terms of the nine parameters in the grouped SPM formulation, as listed in Table~\ref{tab:1}.

\begin{table}[H]
\centering
\caption{Estimated parameters in the grouped SPM formulation and their physical meaning.}
\begin{tabular}{lll}
\hline
\textbf{Symbol} & \textbf{Description} & \textbf{Unit} \\
\hline
$\alpha_{n}$       & Negative electrode time constant           & s \\
$\alpha_{p}$       & Positive electrode time constant           & s \\
$b_n$           & Negative electrode capacity                             & C \\
$b_p$           & Positive electrode capacity                              & C \\
$d_{n}$       &  Negative electrode kinetic grouped parameter         & s·mol$^{1/2}$·m$^{-5/2}$ \\
$d_{p}$       & Positive electrode kinetic grouped parameter          &s·mol$^{1/2}$·m$^{-5/2}$\\
SOC$_n(0)$      & Initial SOC of negative electrode           & - \\
SOC$_p(0)$      & Initial SOC of positive electrode       & - \\
$R_0$ & Reference internal resistance  & $\Omega$ \\
\hline
\end{tabular}
\label{tab:1}
\end{table}

\section{BOLT workflow and implementation}
\begin{figure}[H]
\centering
\makebox[\textwidth][c]{
    \includegraphics[width=1\textwidth]{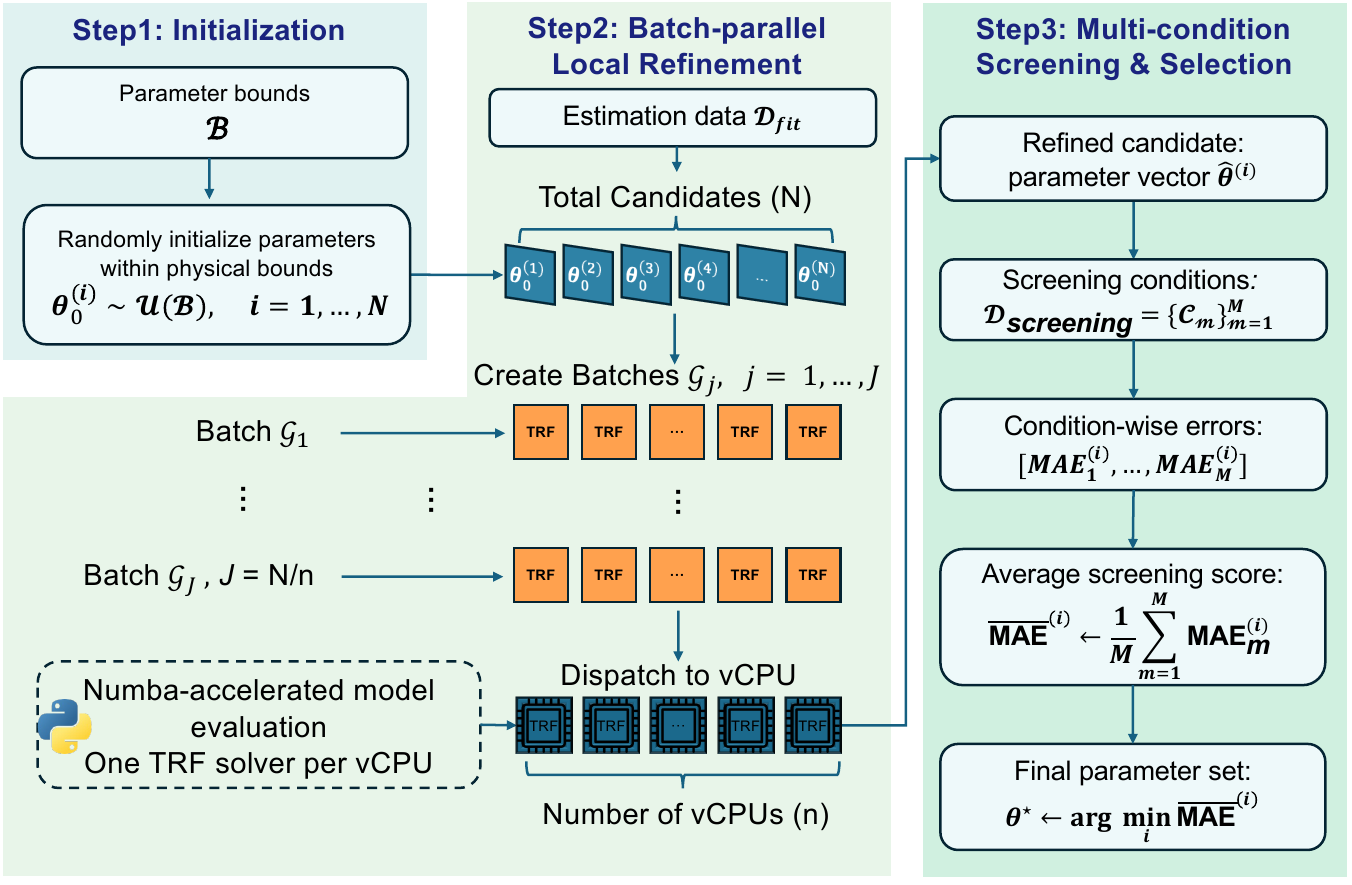}
}
\caption{Workflow of the proposed BOLT framework. Initial parameter vectors are randomly sampled within physically admissible bounds, divided into batches, and refined by parallel TRF solvers using the fitting data set. The refined candidates are then evaluated across multiple operating conditions, and the final parameter vector is selected by minimizing the averaged condition-wise voltage error.}
\label{fig:2}
\end{figure}

The proposed BOLT framework is designed to accelerate physics-based battery parameter estimation while improving robustness across repeated runs. As illustrated in Figure~\ref{fig:2}, an initial set of random parameter vectors is divided into batches and distributed across the available virtual CPUs. Each candidate is refined independently using the \textsc{trf} solver, while the battery model and objective evaluation routines are JIT-accelerated using \texttt{Numba}. After local refinement, all candidate solutions are screened using the mean absolute error averaged over multiple operating conditions, and the lowest-scoring candidate is selected as the final solution. The relationship among the total number of candidates \(N\), the number of cores \(n\), the batch index \(j\), and the screening metric \(\overline{\mathrm{MAE}}\) is summarized in Algorithm~\ref{alg:bolt_ltg}. The implementation details are given below for reproducibility.

\textbf{Step 1 – Initialization.}
We first generate a set of \(N\) candidate parameter vectors
\(\{\boldsymbol{\theta}^{(i)}_{0}\}\) by sampling each component
uniformly within its physically admissible bounds
\(\mathcal{B}\).
The candidates are then divided into \(J=N/n\) equally sized batches,
where \(n\) denotes the number of available virtual CPU cores.
This one-to-one mapping ensures that each core receives exactly one
parameter vector per scheduling cycle and helps avoid load imbalance.

\textbf{Step 2 – Batch-parallel local refinement.}
Each batch \(\mathcal{G}_{j}\) is executed as one parallel scheduling cycle over the available vCPUs. Within each batch, each initial vector \(\boldsymbol{\theta}^{(i)}_{0}\) is dispatched to one vCPU and refined independently using the \textsc{trf} solver on the fitting data set \(\mathcal{D}_{\mathrm{fit}}\). The local optimizer returns a refined parameter vector \(\hat{\boldsymbol{\theta}}^{(i)}\).

\textbf{Step 3 – Multi-condition screening and selection.}
All refined vectors are then assessed on the available current--voltage
data set \(\mathcal{D}_{\mathrm{screening}}
           =\{\mathcal{C}_{1},\dots,\mathcal{C}_{M}\}\).
For each operating condition \(\mathcal{C}_{m}\), a condition-specific
mean absolute error \(\mathrm{MAE}_{m}^{(i)}\) is computed by comparing
the simulated voltage response
\(\mathcal{M}\bigl(\hat{\boldsymbol{\theta}}^{(i)},u_{k,m}\bigr)\)
with the experimental trace \(y_{k,m}^{\mathrm{exp}}\).
These \(M\) errors are averaged to obtain an overall screening score
\(\overline{\mathrm{MAE}}^{(i)}\) for each candidate.
The vector with the lowest averaged score is selected as the final parameter set:
\begin{equation}
\boldsymbol{\theta}^{\star}= \arg\min_{i}\overline{\mathrm{MAE}}^{(i)}.
\label{eq:19}
\end{equation}

Here, the additional operating conditions are used as a candidate-screening criterion for solution selection, rather than as a strict held-out test set. This design reflects the parameter-estimation nature of the problem: unlike supervised machine-learning training, electrochemical model calibration generally benefits from using as much informative current--voltage data as possible to reduce local overfitting and improve parameter identifiability. Therefore, the measured-data results should be interpreted as cross-condition consistency over the available operating-condition matrix, while independent parameter recoverability is examined separately through the synthetic-data tests in Section~\ref{sec:synthetic_recoverability}.

Overall, BOLT combines two complementary ideas: multi-start local refinement and efficient parallel execution. The multi-start strategy increases search diversity and reduces dependence on a single initial guess, while the batch-parallel structure enables simultaneous refinement of multiple candidate solutions. After refinement, multi-condition screening is used to select the most consistent solution among the explored candidates.

In addition to improving repeated-run robustness, BOLT is designed for computational efficiency. All \textsc{trf} instances are executed concurrently on multi-core CPUs. Moreover, the battery model evaluation is JIT-accelerated via \texttt{Numba}, which translates numerically intensive Python code into optimized machine instructions at runtime. The combined effect of parallel execution and JIT-accelerated model evaluation substantially reduces the computational overhead of repeated parameter estimation.

It should be emphasized that JIT acceleration is applied to the battery-model simulation and objective-function evaluation routines only. The internal update mechanisms of TRF, PSO, and GA are not modified. This distinction is important because the speed advantage of BOLT comes from both implementation-level acceleration of repeated model calls and algorithm-level reallocation of the computational budget from population evolution to parallel local refinement and multi-condition candidate screening.

\begin{algorithm}[H]
  \caption{BOLT: Batch-Optimized Local-to-Global Technique}
  \label{alg:bolt_ltg}

  \begin{algorithmic}[1]

    \Require model $\mathcal{M}$,
            parameter bounds $\mathcal{B}$,
            fitting data $\mathcal{D}_{\mathrm{fit}}$,
            screening data set $\mathcal{D}_{\mathrm{screening}}
              =\{\mathcal{C}_{m}\}_{m=1}^{M}$,
            total candidates $N$, vCPU count $n$
    \Statex \hspace{\algorithmicindent} where \(\mathcal{C}_{m}\) denotes the \(m\)-th screening condition and \(M\) is the total number of screening conditions
    \Ensure
        selected parameter vector $\boldsymbol{\theta}^{\star}$

    \Statex \textbf{Step 1 – Initialization}
    \State draw $\{\boldsymbol{\theta}^{(i)}_{0}\}_{i=1}^{N}\sim
           \mathcal{U}(\mathcal{B})$
    \State $J \gets N/n$ \Comment{number of batches}
    \State partition $\{\boldsymbol{\theta}^{(i)}_{0}\}$ into
           batches $\mathcal{G}_{1},\dots,\mathcal{G}_{J}$

    \Statex \textbf{Step 2 – Batch-parallel local optimization}
    \For{batch $j \gets 1$ \textbf{to} $J$}
        \ForAll{$\boldsymbol{\theta}^{(i)}_{0}\in\mathcal{G}_{j}$}
            \State $\hat{\boldsymbol{\theta}}^{(i)} \gets
                   \operatorname{TRF}\bigl(
                   \boldsymbol{\theta}^{(i)}_{0},
                   \mathcal{M},\mathcal{D}_{\mathrm{fit}}\bigr)$
        \EndFor
    \EndFor

    \Statex \textbf{Step 3 – Multi-condition screening and selection}
    \ForAll{refined vector $\hat{\boldsymbol{\theta}}^{(i)}$}
        \For{$m \gets 1$ \textbf{to} $M$}
    \Comment{loop over all screening conditions}
            \State $\displaystyle
              \mathrm{MAE}_{m}^{(i)} \gets
              \frac{1}{|\mathcal{C}_{m}|}\sum_{k}
              \bigl|
                y^{\mathrm{exp}}_{k,m} -
                \mathcal{M}\bigl(\hat{\boldsymbol{\theta}}^{(i)},u_{k,m}\bigr)
              \bigr|$
        \EndFor
        \State $\displaystyle
          \overline{\mathrm{MAE}}^{(i)} \gets
          \frac{1}{M}\sum_{m=1}^{M}\mathrm{MAE}_{m}^{(i)}$
    \EndFor
    \State $\displaystyle
       \boldsymbol{\theta}^{\star} \gets
       \arg\min_{i}\overline{\mathrm{MAE}}^{(i)}$
    \Return $\boldsymbol{\theta}^{\star}$
  \end{algorithmic}
\end{algorithm}

\section{Results and discussion}
In this section, we first describe the comparative experimental settings and then evaluate the proposed BOLT framework against the baseline PSO and GA methods. The discussion focuses on three aspects that are central to practical electrochemical model calibration: estimation accuracy, computational efficiency, and repeated-run stability.

\subsection{Comparative experimental settings}\label{sec4}

To evaluate the proposed BOLT method, we compare it with two widely used optimization approaches for electrochemical model parameter estimation: PSO and GA. All algorithms were implemented in Python and executed on a Linux-based computation platform equipped with an Intel(R) Xeon(R) Gold 5120 CPU @ 2.20 GHz, 16 virtual CPUs, and 32 GB of memory. PSO and GA were implemented using the corresponding modules from the \texttt{scikit-opt} library. The TRF solver used in BOLT was implemented via the \texttt{least\_squares} function from \texttt{scipy.optimize}, with \texttt{method='trf'}. This solver is based on a trust-region subproblem with bound constraints, as formulated by Branch et al.~\cite{branch1999subspace}.

\begin{table}[htbp]
\centering
\caption{Hyperparameter settings for BOLT, PSO, and GA.}
\label{tab:hyperparams}
\begin{tabular}{@{}l p{0.7\linewidth}@{}}
\toprule
\textbf{Algorithm} & \textbf{Hyperparameters} \\
\midrule
\textbf{BOLT (TRF)} & \texttt{method='trf'}, \texttt{ftol=1e-8}, \texttt{xtol=1e-8}, \texttt{gtol=1e-8}, \texttt{max\_nfev=50000}, \texttt{n=16} \\
\textbf{PSO} & \texttt{pop=100}, \texttt{max\_iter=500}, \texttt{w=0.8}, \texttt{c1=0.5}, \texttt{c2=0.5}, \texttt{n\_processes=16} \\
\textbf{GA} & \texttt{size\_pop=100}, \texttt{max\_iter=500}, \texttt{prob\_mut=0.001}, \texttt{precision=1e-8}, \texttt{n\_processes=16} \\
\bottomrule
\end{tabular}
\end{table}

The TRF tolerances in Table~\ref{tab:hyperparams} follow the default tolerance level used by \texttt{scipy.optimize.least\_squares}. No operating-condition-specific tuning was applied to the optimizer settings. The large value of \texttt{max\_nfev} was used only as a safety upper bound to avoid premature termination by a fixed evaluation cap. Similarly, the batch configuration was determined by the available 16 virtual CPU cores; therefore, candidate numbers were selected as multiples of 16 so that the number of batches was an integer and the available parallel resources could be fully utilized.

To improve fairness, PSO and GA were evaluated under both two-condition and five-condition settings. In the two-condition setting, only the 0.5C and 1C profiles were used during optimization. This choice was not arbitrary: our previous comparative study showed that, among several combinations of operating profiles, the 0.5C and 1C pair provided a favorable balance between estimation accuracy and computational efficiency for electrochemical battery model parameter estimation \cite{guo2026optimizing}. In the five-condition setting, all five operating conditions (0.2C, 0.333C, 0.5C, 1C, and DST) were used during optimization. In addition, for PSO and GA, execution time was reported both without JIT acceleration and with JIT acceleration using the same Numba-accelerated battery model evaluation routines. In this way, the comparison better distinguishes algorithmic differences from implementation-level acceleration. The corresponding hyperparameter settings are summarized in Table~\ref{tab:hyperparams}, and the parameter ranges of the grouped SPM are given in Table~\ref{tab:parameter_ranges_grouped}~\cite{guo2025control}.

\begin{table}[htbp]
\centering
\caption{Parameter ranges.}
\label{tab:parameter_ranges_grouped}
\begin{tabular}{lll}
\toprule
\textbf{Parameter} & \textbf{Range} & \textbf{Unit} \\
\midrule
$\alpha_n$ & [625, 7692] & s \\
$\alpha_p$ & [1.587, 2500] & s \\
$d_n$ & [$5.7 \times 10^{-5}$, $7.8 \times 10^{-4}$] & s$\cdot$mol$^{1/2}\cdot$m$^{-5/2}$ \\
$d_p$ & [$7.9 \times 10^{-5}$, $1.0 \times 10^{-3}$] & s$\cdot$mol$^{1/2}\cdot$m$^{-5/2}$ \\
$b_n$ & [$8352$, $12528$] & C \\
$b_p$ & [$8352$, $12528$] & C \\
SOC$_n(0)$ & [0.8, 1.0] & -- \\
SOC$_p(0)$ & [0, 0.2] & -- \\
$R_0$ & [0, 0.05] & $\Omega$ \\
\bottomrule
\end{tabular}
\end{table}

\begin{figure}[H]
\centering
\makebox[\textwidth][c]{
    \includegraphics[width=1.1\textwidth]{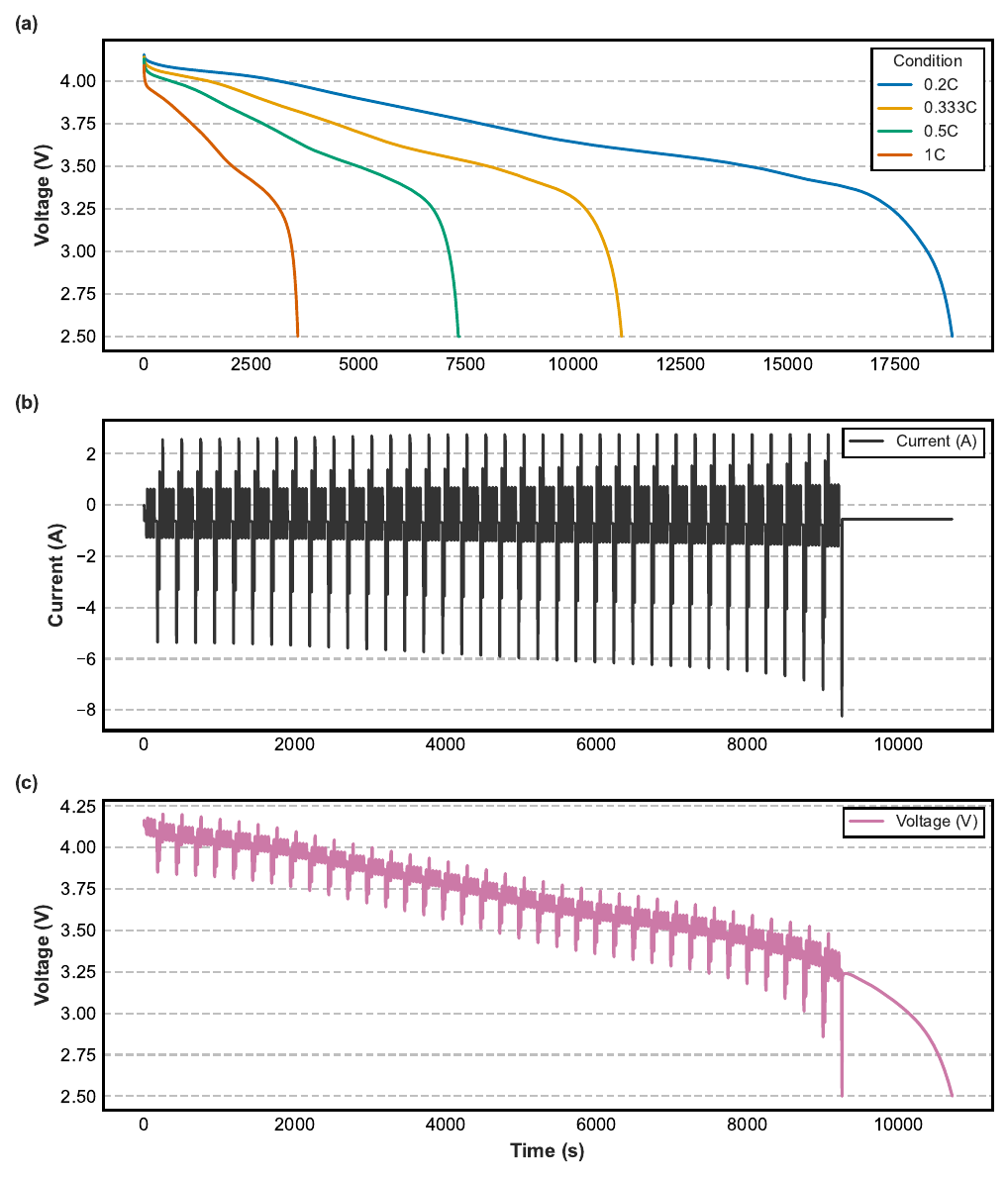}
}
\caption{Operating conditions considered in this study. (a) Measured terminal-voltage curves under four constant-current discharge conditions: 0.2C, 0.333C, 0.5C, and 1C. (b) Applied current profile of the DST. (c) Measured terminal-voltage response under the DST condition.}
\label{fig:3}
\end{figure}

In this study, a commercially available 18650 lithium-ion cell with an NMC cathode, a graphite anode, and a nominal capacity of 2.9 Ah was used. The grouped SPM described in Section~2 was used for all parameter estimation experiments. The test conditions included constant-current discharge at 0.2C, 0.333C, 0.5C, and 1C, together with a DST profile (see Figure~\ref{fig:3}). For BOLT, the 0.5C and 1C profiles were used in Step 2 for local parameter refinement, while the five-condition set was used in Step 3 as a candidate-screening criterion. This two-condition local-refinement design is consistent with the profile selection strategy identified in our previous study \cite{guo2026optimizing}, which showed that the combination of the 0.5C and 1C profiles provides a favorable trade-off between estimation accuracy and computational efficiency. For PSO and GA, both two-condition and five-condition optimization settings were examined explicitly, as described above. Each algorithm was independently run 50 times, and all reported results were computed from these 50 repeated runs. This repeated-run evaluation design makes it possible to assess not only estimation accuracy, but also computational efficiency and repeated-run stability, all of which are critical for practical electrochemical model calibration.

\subsection{Effect of Total Candidate Number $N$}

\begin{figure}[H]
\centering
\makebox[\textwidth][c]{
    \includegraphics[width=1\textwidth]{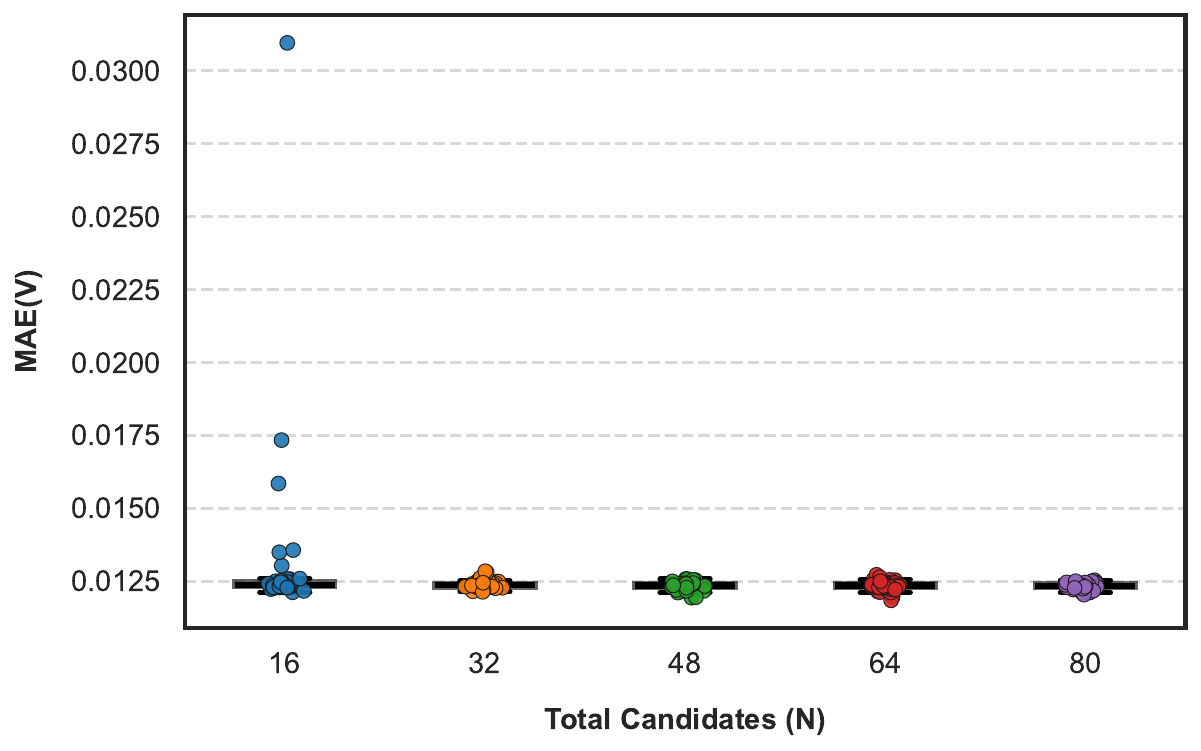}
}
\caption{Effect of the total candidate number $N$ on the MAE distribution of the BOLT method over 50 repeated runs.}
\label{fig:4}
\end{figure}

We first examine the influence of the total candidate number $N$ on the performance of the proposed BOLT framework. Figure~\ref{fig:4} presents the MAE distributions obtained with $N=16$, $32$, $48$, $64$, and $80$, while Table~\ref{tab:3} reports the corresponding summary statistics. These values were selected as integer multiples of the 16 available virtual CPUs, corresponding to 1, 2, 3, 4, and 5 batches, respectively. This configuration enables efficient batch-parallel execution of the candidate solutions.

As shown in Figure~\ref{fig:4}, the BOLT results with $N=16$ exhibit visibly higher dispersion than the other settings, including several clear high-error outliers, whereas the distributions for $N=32$, $48$, $64$, and $80$ are much more concentrated around approximately 0.012--0.013 V. This indicates that a single batch of candidates may not provide sufficient search diversity to consistently identify a low-error solution. Once the number of candidates is increased to $N=32$, the distribution contracts markedly, and further increases in $N$ mainly reduce the already small residual variation rather than producing substantial additional gains.

This trend is consistent with the numerical results in Table~\ref{tab:3}. BOLT(16) yields \(13.0 \pm 2.7\) mV, whereas BOLT(32) already reaches \(12.4 \pm 0.1\) mV. Further increasing $N$ to 48, 64, and 80 only reduces the average error slightly to approximately \(12.3 \pm 0.1\) mV. Therefore, increasing $N$ improves the probability of identifying a lower-error solution within the explored candidate pool, but this benefit gradually saturates once a sufficient level of search diversity has been achieved.

From an efficiency perspective, the computational cost continues to increase with $N$ because more local refinement tasks must be executed. However, even at $N=80$, the total runtime remains only \(22.24 \pm 2.17\) s, which is still very small compared with the population-based baselines in Table~\ref{tab:3}. Taken together, these results show that the total candidate number controls the trade-off between search diversity and computational cost in BOLT. In the present study, $N=32$ provides the most favorable balance among estimation accuracy, computational efficiency, and repeated-run stability. It is therefore selected as the representative BOLT configuration for the subsequent comparisons with PSO and GA.

\subsection{Synthetic parameter recoverability and noise robustness}
\label{sec:synthetic_recoverability}

The preceding real-data comparison demonstrates the practical advantage of BOLT in reproducing measured voltage responses across multiple operating conditions. To isolate the behavior of the optimization workflow itself, an additional synthetic-data experiment was conducted to examine whether BOLT can recover a known parameter vector in the grouped SPM formulation under controlled conditions.This experiment is therefore interpreted as a parameter recoverability and noise-robustness test, rather than as an independent real-data validation.

A reference parameter vector in the grouped SPM formulation was first used to generate synthetic voltage responses with the grouped SPM. Five operating profiles were considered, consistent with the main experimental setting: 0.2C, 0.333C, 0.5C, 1C, and DST. For the constant-current cases, synthetic current profiles were generated until the terminal voltage reached the cut-off voltage of 2.5 V, while the DST case used the measured current profile. In each BOLT run, the 0.5C and 1C profiles were used for local parameter refinement, while all five profiles were used for multi-condition consistency evaluation. Each BOLT run contained \(N=32\) randomly initialized local optimizations, and the whole procedure was repeated 30 times.

To examine robustness against voltage measurement noise, zero-mean white Gaussian noise with standard deviations of 1, 2, and 3 mV was added to the synthetic voltage signals. The noisy voltage signals were then used as the input for parameter estimation. After parameter estimation, the reconstructed voltage was compared with both the fitted voltage signal and the underlying clean synthetic voltage. Here, the fitted voltage denotes the clean synthetic voltage in the noiseless case and the noisy voltage in the noise-injected cases. For consistency with the main comparison, all voltage reconstruction errors in this subsection are reported using mean absolute error (MAE).

For a target voltage signal \(y^{\mathrm{tar}}_{k,m}\), the condition-wise voltage MAE is defined as
\begin{equation}
\mathrm{MAE}^{\mathrm{tar}}_{m}
=
\frac{1}{|\mathcal{C}_{m}|}
\sum_{k\in\mathcal{C}_{m}}
\left|
y^{\mathrm{tar}}_{k,m}
-
\mathcal{M}\left(\hat{\boldsymbol{\theta}},u_{k,m}\right)
\right|,
\label{eq:synthetic_voltage_mae}
\end{equation}
where \(\mathcal{C}_{m}\) denotes the \(m\)-th operating profile, \(u_{k,m}\) is the applied current, and \(\mathcal{M}\left(\hat{\boldsymbol{\theta}},u_{k,m}\right)\) is the voltage simulated using the estimated parameter vector. The average voltage MAE over the five operating profiles is then computed as
\begin{equation}
\overline{\mathrm{MAE}}^{\mathrm{tar}}
=
\frac{1}{M}
\sum_{m=1}^{M}
\mathrm{MAE}^{\mathrm{tar}}_{m},
\quad M=5.
\label{eq:synthetic_avg_mae}
\end{equation}

The parameter recovery error was evaluated using absolute relative error (ARE), because the nine estimated parameters have different units and numerical scales. For the \(j\)-th parameter, ARE is defined as
\begin{equation}
\mathrm{ARE}_{j}
=
\left|
\frac{\hat{\theta}_{j}-\theta^{\mathrm{ref}}_{j}}
{\theta^{\mathrm{ref}}_{j}}
\right|
\times 100\%,
\label{eq:param_are}
\end{equation}
where \(\theta^{\mathrm{ref}}_{j}\) is the reference value and \(\hat{\theta}_{j}\) is the corresponding estimated value. The mean parameter ARE is computed as
\begin{equation}
\overline{e}_{\theta}
=
\frac{1}{n_{\theta}}
\sum_{j=1}^{n_{\theta}}
\mathrm{ARE}_{j},
\quad n_{\theta}=9.
\label{eq:param_mean_are}
\end{equation}

\begin{table}[H]
\centering
\caption{Synthetic recoverability and noise robustness of BOLT(32) over the five operating profiles. Values are reported as mean \(\pm\) standard deviation over 30 independent BOLT runs.}
\label{tab:synthetic_recoverability_noise}
\renewcommand{\arraystretch}{0.92}
\setlength{\tabcolsep}{3.2pt}
\footnotesize
\begin{tabular}{@{}c c c c@{}}
\toprule
\textbf{Noise} &
\shortstack{\textbf{Fitted}\\\textbf{MAE}} &
\shortstack{\textbf{Clean}\\\textbf{MAE}} &
\shortstack{\textbf{Mean param.}\\\textbf{ARE}} \\
\textbf{(mV)} &
\textbf{(mV)} &
\textbf{(mV)} &
\textbf{(\%)} \\
\midrule
0 & \(<10^{-10}\) & \(<10^{-10}\) & \(<10^{-9}\) \\
1 & \(0.799\pm{<}10^{-3}\) & \(0.023\pm{<}10^{-3}\) & \(0.218\pm{<}10^{-3}\) \\
2 & \(1.597\pm{<}10^{-3}\) & \(0.028\pm{<}10^{-3}\) & \(0.424\pm{<}10^{-3}\) \\
3 & \(2.398\pm{<}10^{-3}\) & \(0.103\pm{<}10^{-3}\) & \(0.507\pm{<}10^{-3}\) \\
\bottomrule
\end{tabular}

\vspace{2pt}
\begin{minipage}{0.82\textwidth}
\scriptsize
ARE denotes absolute relative error. Fitted MAE is computed against the voltage used during estimation, while clean MAE is computed against the underlying noiseless synthetic voltage. Values in the noiseless case are below the displayed precision.
\end{minipage}
\end{table}

\begin{figure}[H]
\centering
\makebox[\textwidth][c]{
    \includegraphics[width=0.80\textwidth]{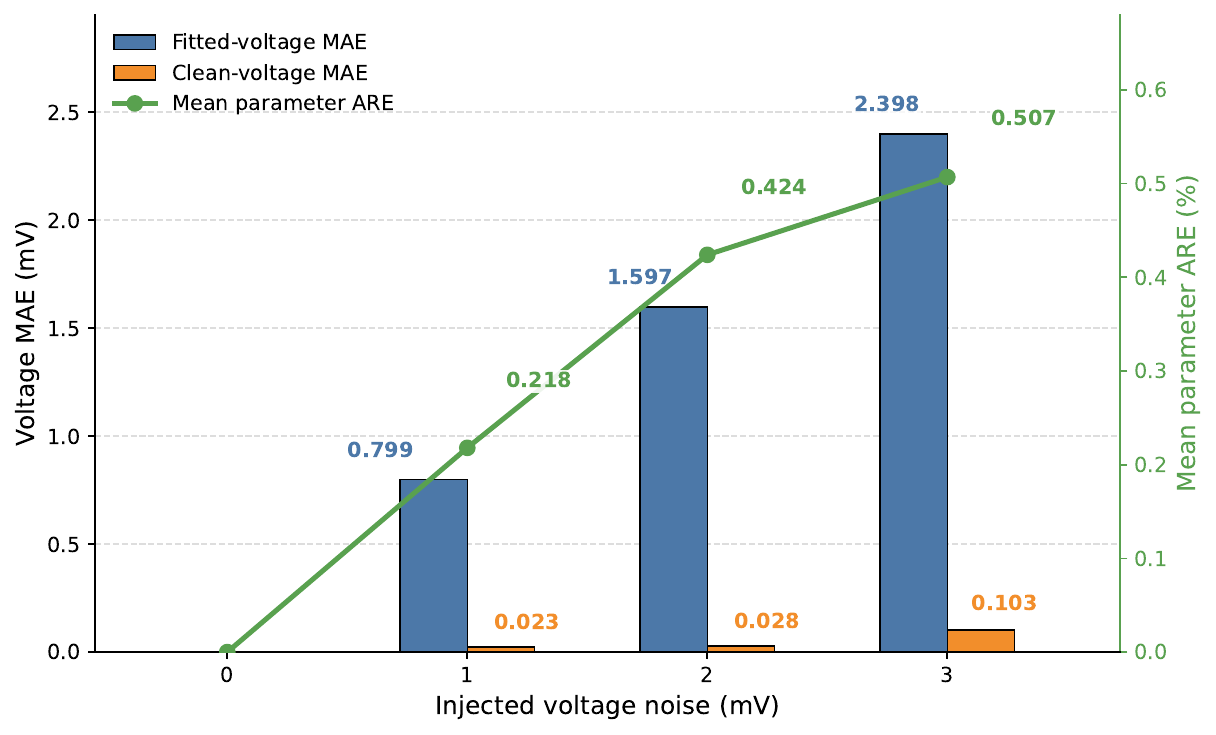}
}
\caption{Synthetic recoverability and noise robustness of BOLT(32) over the five operating profiles. The bars show the voltage MAE with respect to the fitted voltage and the underlying clean synthetic voltage, while the line indicates the mean parameter ARE over 30 independent BOLT runs.}
\label{fig:synthetic_mae_noise_robustness}
\end{figure}

As shown in Table~\ref{tab:synthetic_recoverability_noise} and Figure~\ref{fig:synthetic_mae_noise_robustness}, BOLT recovered the reference parameter vector with numerical precision in the noiseless synthetic-data setting. Both the voltage MAE and the mean parameter ARE were essentially zero at the displayed precision. This result confirms the numerical self-consistency of the proposed workflow: when the voltage data are generated by the same grouped SPM and no measurement noise is present, BOLT can accurately recover the given parameter vector in the grouped SPM formulation using only the 0.5C and 1C profiles for local refinement.

When voltage noise was added, the fitted-voltage MAE increased approximately in proportion to the injected noise level. The fitted-voltage MAE values were 0.799, 1.597, and 2.398 mV under 1, 2, and 3 mV noise, respectively. This trend is consistent with the expected behavior of zero-mean Gaussian noise, because the mean absolute deviation of a Gaussian noise signal is lower than its standard deviation. More importantly, when the reconstructed voltage was compared with the underlying clean synthetic voltage, the MAE remained far below the injected noise level. The clean-voltage MAE was only 0.023 mV under 1 mV noise, 0.028 mV under 2 mV noise, and 0.103 mV under 3 mV noise. This indicates that BOLT does not simply overfit the added white noise; instead, it can still recover parameter sets that closely reproduce the underlying clean model response.

The recovered parameter values are further summarized in Table~\ref{tab:synthetic_parameter_recovery}. The estimates are reported as mean \(\pm\) standard deviation over 30 independent BOLT runs. In the noiseless case, all estimated parameters were identical to the reference values at the displayed precision, and the standard deviations were also close to numerical precision. Under voltage-noise perturbations, the estimated parameters remained close to the reference values. Even under 3 mV voltage noise, the mean parameter ARE was only 0.507\%.  Across all tested noise levels, the mean parameter ARE remained below \(0.6\%\), which supports the parameter-recovery robustness of BOLT under millivolt-level voltage perturbations. The largest deviations appeared mainly in the positive-electrode kinetic grouped parameter \(d_p\), whereas the capacity-related parameters, initial stoichiometries, and ohmic resistance remained accurately recovered.

\begin{table}[H]
\centering
\caption{Recovered parameters in the grouped SPM formulation under noiseless and noisy synthetic-data settings. Values are reported as mean \(\pm\) standard deviation over 30 BOLT runs.}
\label{tab:synthetic_parameter_recovery}
\renewcommand{\arraystretch}{0.88}
\setlength{\tabcolsep}{3.0pt}
\scriptsize
\begin{tabularx}{\textwidth}{
l
c
c
>{\centering\arraybackslash}X
>{\centering\arraybackslash}X
}
\toprule
\textbf{Parameter} &
\textbf{Noise} &
\textbf{Reference} &
\textbf{Recovered estimate} &
\textbf{ARE (\%)} \\
&
\textbf{(mV)} &
&
\textbf{mean \(\pm\) std} &
\textbf{mean \(\pm\) std} \\
\midrule

\multirow{4}{*}{\(\alpha_n\)}
& 0 & \multirow{4}{*}{2746.771743} & \(\theta^{\mathrm{ref}}\) & \(<10^{-9}\) \\
& 1 &  & \(2746.401268 \pm 1.40{\times}10^{-5}\) & \(0.0135 \pm 5.09{\times}10^{-7}\) \\
& 2 &  & \(2747.309383 \pm 4.51{\times}10^{-5}\) & \(0.0196 \pm 1.64{\times}10^{-6}\) \\
& 3 &  & \(2746.065877 \pm 6.33{\times}10^{-5}\) & \(0.0257 \pm 2.30{\times}10^{-6}\) \\

\multirow{4}{*}{\(\alpha_p\)}
& 0 & \multirow{4}{*}{1352.591585} & \(\theta^{\mathrm{ref}}\) & \(<10^{-9}\) \\
& 1 &  & \(1346.799011 \pm 1.28{\times}10^{-3}\) & \(0.4283 \pm 9.46{\times}10^{-5}\) \\
& 2 &  & \(1338.464239 \pm 2.23{\times}10^{-3}\) & \(1.0445 \pm 1.65{\times}10^{-4}\) \\
& 3 &  & \(1363.279512 \pm 3.32{\times}10^{-3}\) & \(0.7902 \pm 2.45{\times}10^{-4}\) \\

\multirow{4}{*}{\(b_n\)}
& 0 & \multirow{4}{*}{11033.60104} & \(\theta^{\mathrm{ref}}\) & \(<10^{-9}\) \\
& 1 &  & \(11031.95155 \pm 2.85{\times}10^{-5}\) & \(0.0150 \pm 2.59{\times}10^{-7}\) \\
& 2 &  & \(11034.94807 \pm 4.53{\times}10^{-5}\) & \(0.0122 \pm 4.11{\times}10^{-7}\) \\
& 3 &  & \(11026.84710 \pm 1.54{\times}10^{-4}\) & \(0.0612 \pm 1.39{\times}10^{-6}\) \\

\multirow{4}{*}{\(b_p\)}
& 0 & \multirow{4}{*}{10965.02498} & \(\theta^{\mathrm{ref}}\) & \(<10^{-9}\) \\
& 1 &  & \(10963.74726 \pm 3.04{\times}10^{-5}\) & \(0.0117 \pm 2.77{\times}10^{-7}\) \\
& 2 &  & \(10965.17623 \pm 5.21{\times}10^{-5}\) & \(0.0014 \pm 4.75{\times}10^{-7}\) \\
& 3 &  & \(10956.73912 \pm 1.38{\times}10^{-4}\) & \(0.0756 \pm 1.26{\times}10^{-6}\) \\

\multirow{4}{*}{\(d_n\)}
& 0 & \multirow{4}{*}{\(7.6000{\times}10^{-5}\)} & \(\theta^{\mathrm{ref}}\) & \(<10^{-9}\) \\
& 1 &  & \((7.6190{\times}10^{-5}) \pm 2.07{\times}10^{-11}\) & \(0.2504 \pm 2.72{\times}10^{-5}\) \\
& 2 &  & \((7.6014{\times}10^{-5}) \pm 3.79{\times}10^{-11}\) & \(0.0188 \pm 4.98{\times}10^{-5}\) \\
& 3 &  & \((7.7036{\times}10^{-5}) \pm 5.10{\times}10^{-11}\) & \(1.3627 \pm 6.72{\times}10^{-5}\) \\

\multirow{4}{*}{\(d_p\)}
& 0 & \multirow{4}{*}{\(6.2200{\times}10^{-4}\)} & \(\theta^{\mathrm{ref}}\) & \(<10^{-9}\) \\
& 1 &  & \((6.2902{\times}10^{-4}) \pm 1.13{\times}10^{-10}\) & \(1.1291 \pm 1.81{\times}10^{-5}\) \\
& 2 &  & \((6.0777{\times}10^{-4}) \pm 2.34{\times}10^{-10}\) & \(2.2881 \pm 3.77{\times}10^{-5}\) \\
& 3 &  & \((6.3489{\times}10^{-4}) \pm 8.56{\times}10^{-10}\) & \(2.0726 \pm 1.38{\times}10^{-4}\) \\

\multirow{4}{*}{SOC\(_n(0)\)}
& 0 & \multirow{4}{*}{0.930129} & \(\theta^{\mathrm{ref}}\) & \(<10^{-9}\) \\
& 1 &  & \(0.930263 \pm 2.61{\times}10^{-9}\) & \(0.0144 \pm 2.80{\times}10^{-7}\) \\
& 2 &  & \(0.930016 \pm 4.19{\times}10^{-9}\) & \(0.0121 \pm 4.51{\times}10^{-7}\) \\
& 3 &  & \(0.930691 \pm 1.32{\times}10^{-8}\) & \(0.0604 \pm 1.42{\times}10^{-6}\) \\

\multirow{4}{*}{SOC\(_p(0)\)}
& 0 & \multirow{4}{*}{0.036550} & \(\theta^{\mathrm{ref}}\) & \(<10^{-9}\) \\
& 1 &  & \(0.036583 \pm 8.54{\times}10^{-9}\) & \(0.0905 \pm 2.34{\times}10^{-5}\) \\
& 2 &  & \(0.036579 \pm 1.49{\times}10^{-8}\) & \(0.0793 \pm 4.07{\times}10^{-5}\) \\
& 3 &  & \(0.036578 \pm 2.21{\times}10^{-8}\) & \(0.0768 \pm 6.05{\times}10^{-5}\) \\

\multirow{4}{*}{\(R_0\)}
& 0 & \multirow{4}{*}{0.032005} & \(\theta^{\mathrm{ref}}\) & \(<10^{-9}\) \\
& 1 &  & \(0.032001 \pm 7.54{\times}10^{-9}\) & \(0.0119 \pm 2.36{\times}10^{-5}\) \\
& 2 &  & \(0.032114 \pm 1.31{\times}10^{-8}\) & \(0.3408 \pm 4.09{\times}10^{-5}\) \\
& 3 &  & \(0.031992 \pm 2.01{\times}10^{-8}\) & \(0.0407 \pm 6.28{\times}10^{-5}\) \\

\bottomrule
\end{tabularx}

\vspace{2pt}
\begin{minipage}{0.96\textwidth}
\scriptsize
ARE denotes absolute relative error. For the 0 mV case, \(\theta^{\mathrm{ref}}\) indicates that the recovered estimate is identical to the reference value at the displayed precision. The corresponding ARE values are below numerical precision and are therefore reported as \(<10^{-9}\%\).
\end{minipage}
\end{table}

Overall, the synthetic-data tests complement the real-data comparisons by separating the recoverability of the optimization workflow from model-structure mismatch and experimental uncertainty. The noiseless case confirms that the BOLT implementation can recover the  known parameter vector when the data are internally consistent with the model. The noise-injected cases further show that both the reconstructed clean-voltage response and the recovered parameters remain stable under realistic millivolt-level voltage perturbations. Therefore, the larger errors observed in measured-data calibration are more likely associated with experimental noise, model-structure mismatch, and condition-to-condition inconsistencies, rather than with a failure of the BOLT optimization procedure itself.

\subsection{Overall Comparison with PSO and GA}

Table~\ref{tab:3} summarizes the statistical performance of the proposed BOLT method across different total candidate numbers, together with PSO and GA under both two-condition and five-condition settings. In this table, \textbf{BOLT($N$)} denotes the BOLT variant executed with a total of $N$ candidate solutions. \textbf{Avg\_MAE} denotes the average mean absolute error over the five operating conditions (0.2C, 0.333C, 0.5C, 1C, and DST), aggregated over 50 repeated runs. \textbf{Battery model calls} denote the total number of electrochemical model evaluations. For PSO and GA, the total number of model calls is $100 \times 500 \times 2 = 100{,}000$ in the two-condition setting and $100 \times 500 \times 5 = 250{,}000$ in the five-condition setting. \textbf{Time} denotes the wall-clock runtime for one complete optimization run, and \textbf{Peak memory} denotes the maximum memory usage observed during one complete optimization run.

To further assess computational resource requirements, per-run peak memory usage was recorded during the optimization stage. Memory profiling was implemented in Python using the \texttt{psutil} library. For each profiling run, the resident set size (RSS) was sampled at 0.2-s intervals for the full process tree, including the main Python process and all parallel worker processes. The reported peak memory corresponds to the maximum total RSS observed during one complete optimization run. Before memory profiling, the Numba-compiled battery model was warmed up once to avoid counting one-time just-in-time compilation overhead as part of the optimization memory cost. For BOLT, memory results were obtained over 50 repeated profiling runs for each total candidate number. For PSO and GA, memory usage was also measured over 50 repeated runs under both two-condition and five-condition settings.

\begin{table}[H]
\centering
\scriptsize
\caption{Statistical results for BOLT with different total candidate numbers, compared with PSO and GA under two-condition and five-condition settings. Accuracy, per-run runtime, and per-run peak memory results are reported as mean $\pm$ std over 50 repeated runs. Peak memory denotes the peak process-tree RSS memory usage during one complete optimization run.}
\label{tab:3}
\renewcommand{\arraystretch}{1.08}
\setlength{\tabcolsep}{2.5pt}
\begin{tabular}{@{}l c c l c@{}}
\toprule
\textbf{Method} &
\textbf{Avg\_MAE (mV)} &
\textbf{Model calls} &
\textbf{Time (s)} &
\textbf{Peak memory (MB)} \\
\midrule
BOLT(16) & \(13.0 \pm 2.7\) & \(11101 \pm 2162\) & \(5.37 \pm 1.14\) & \(4564.8 \pm 8.8\) \\
BOLT(32) & \(12.4 \pm 0.1\) & \(20636 \pm 3081\) & \(8.97 \pm 1.20\) & \(4561.6 \pm 3.8\) \\
BOLT(48) & \(12.3 \pm 0.1\) & \(31660 \pm 3735\) & \(13.49 \pm 1.92\) & \(4554.3 \pm 4.4\) \\
BOLT(64) & \(12.3 \pm 0.1\) & \(43188 \pm 5240\) & \(17.88 \pm 1.83\) & \(4517.1 \pm 8.1\) \\
BOLT(80) & \(12.3 \pm 0.1\) & \(53424 \pm 5321\) & \(22.24 \pm 2.17\) & \(4539.6 \pm 8.7\) \\
\midrule
PSO (2 cond.) & \(15.4 \pm 1.8\) & \(100000 \pm 0\) &
\shortstack[l]{without JIT: \(3506.84 \pm 56.87\) \\ JIT: \(255.53 \pm 0.42\)} &
\(3930.4 \pm 0.8\) \\
\midrule
PSO (5 cond.) & \(14.1 \pm 1.2\) & \(250000 \pm 0\) &
\shortstack[l]{without JIT: \(32291.01 \pm 23.50\) \\ JIT: \(858.88 \pm 2.48\)} &
\(3950.2 \pm 2.4\) \\
\midrule
GA (2 cond.) & \(14.5 \pm 3.3\) & \(100000 \pm 0\) &
\shortstack[l]{without JIT: \(3594.76 \pm 17.90\) \\ JIT: \(269.28 \pm 0.60\)} &
\(3917.7 \pm 1.0\) \\
\midrule
GA (5 cond.) & \(13.3 \pm 1.0\) & \(250000 \pm 0\) &
\shortstack[l]{without JIT: \(32782.78 \pm 21.46\) \\ JIT: \(867.92 \pm 1.56\)} &
\(3952.8 \pm 1.3\) \\
\bottomrule
\end{tabular}

\vspace{2pt}
\begin{minipage}{0.98\linewidth}
\scriptsize
BOLT($N$): BOLT with total candidate number $N$.
PSO/GA (2 cond.) use 0.5C and 1C during optimization.
PSO/GA (5 cond.) use 0.2C, 0.333C, 0.5C, 1C, and DST during optimization.
Avg\_MAE denotes the average mean absolute error over five operating conditions.
Model calls denote total electrochemical model evaluations.
Peak memory denotes the peak resident set size (RSS) of the full process tree, including the main process and all parallel worker processes, measured during the optimization stage.
\end{minipage}
\end{table}

In terms of per-run computational efficiency, BOLT substantially reduced both the number of battery model evaluations and the wall-clock runtime compared with PSO and GA. Even when the total candidate number increased from 16 to 80, BOLT required fewer model calls than PSO and GA and completed the optimization within tens of seconds under the JIT-accelerated implementation. For example, BOLT(32) required only \(20636 \pm 3081\) model calls and \(8.97 \pm 1.20\) s per run, whereas PSO and GA required 100,000--250,000 model calls and substantially longer runtime under the corresponding settings.

The peak memory usage of BOLT was slightly higher than that of PSO and GA, remaining around 4.5 GB across different total candidate numbers. For instance, BOLT(32) used \(4561.6 \pm 3.8\) MB, compared with \(3930.4 \pm 0.8\) MB for PSO (2 cond.), \(3950.2 \pm 2.4\) MB for PSO (5 cond.), \(3917.7 \pm 1.0\) MB for GA (2 cond.), and \(3952.8 \pm 1.3\) MB for GA (5 cond.). This moderate increase is mainly because BOLT performs multiple independent local least-squares optimizations in parallel, and each worker process maintains solver-specific intermediate arrays. By contrast, PSO and GA mainly evaluate population candidates through the objective function, leading to slightly lower per-worker memory usage. For PSO and GA, increasing the number of optimization conditions from two to five led to only a modest increase in peak memory usage, indicating that memory consumption was affected by the larger objective-function workload but remained primarily governed by the number of concurrent worker processes. Importantly, the peak memory usage of BOLT remained nearly unchanged when the total candidate number increased from 16 to 80, because BOLT limits the number of concurrently executed workers to 16 and processes additional candidates in batches. Thus, increasing the total candidate number mainly increased runtime rather than peak memory consumption. Overall, BOLT introduces a moderate increase in peak memory usage but achieves substantially lower runtime and fewer model evaluations than PSO and GA. In addition, the tested configurations were evaluated on a workstation with 32 GB of RAM, and the measured peak memory usage remained well below the available memory capacity. Therefore, memory consumption was not the computational bottleneck in the present experiments.

\begin{figure}[H]
\centering
\makebox[\textwidth][c]{
    \includegraphics[width=1.3\textwidth]{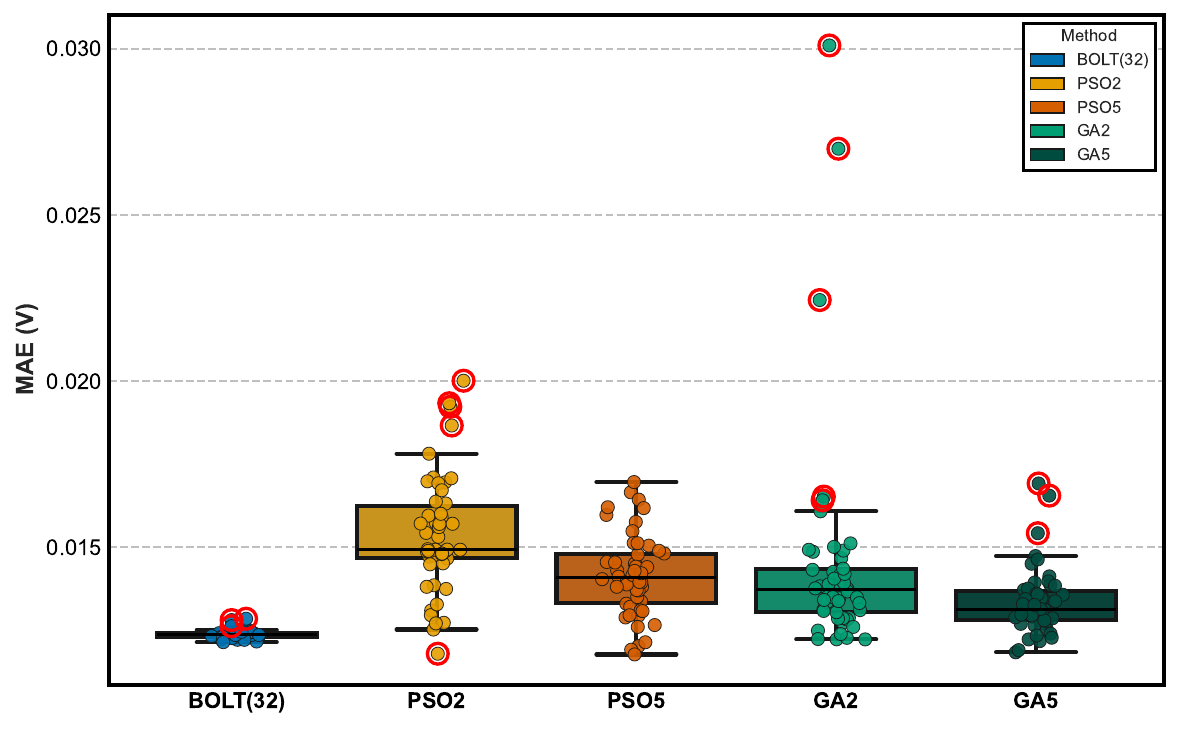}
}
\caption{Overall MAE distributions of BOLT(32), PSO2, PSO5, GA2, and GA5 over 50 repeated runs. Here, PSO2/GA2 denote the two-condition setting (0.5C and 1C), whereas PSO5/GA5 denote the five-condition setting (0.2C, 0.333C, 0.5C, 1C, and DST).}
\label{fig:5}
\end{figure}

Figure~\ref{fig:5} compares the overall MAE distributions of the five methods: BOLT(32), PSO (2 cond.), PSO (5 cond.), GA (2 cond.), and GA (5 cond.), denoted in the figure legend as BOLT(32), PSO2, PSO5, GA2, and GA5, respectively. The box plot shows that BOLT(32) has the lowest central error level and the tightest overall distribution, indicating stronger repeated-run stability than the compared PSO and GA settings. In contrast, PSO2 and GA2 display substantially broader spreads and more pronounced high-error outliers. PSO5 and GA5 improve relative to their two-condition counterparts, but still remain above BOLT(32) in both median error and distribution width.

Three observations are particularly important. First, increasing the number of operating conditions during optimization improves the performance of both PSO and GA. PSO (5 cond.) outperforms PSO (2 cond.), and GA (5 cond.) similarly outperforms GA (2 cond.), indicating that broader condition coverage improves the overall fitting quality across the available operating-condition set. However, this gain is achieved at the cost of a 2.5-fold increase in battery model calls, from 100,000 to 250,000.

Second, JIT acceleration significantly reduces the runtime of PSO and GA as well. For example, PSO (2 cond.) is reduced from \(3506.84 \pm 56.87\) s to \(255.53 \pm 0.42\) s, and GA (2 cond.) is reduced from \(3594.76 \pm 17.90\) s to \(269.28 \pm 0.60\) s. Under the five-condition setting, JIT acceleration shortens the runtime of PSO and GA from more than 32,000 s to approximately 860 s. Nevertheless, BOLT(32) still completes in only \(8.97 \pm 1.20\) s per run, making it roughly 28--30 times faster than the JIT-accelerated two-condition PSO/GA baselines and nearly two orders of magnitude faster than the JIT-accelerated five-condition settings. These results indicate that the efficiency advantage of BOLT cannot be attributed solely to implementation-level acceleration, but also to its lower number of model evaluations and its more efficient local-to-global calibration workflow.

The speedup obtained by JIT acceleration is not strictly linear with the number of operating conditions. The total runtime contains fixed Python overhead, optimization-control overhead, and model-evaluation-dominated computation. When more operating conditions are included, a larger fraction of the workload is spent on repeated battery-model simulation and objective-function evaluation, where JIT acceleration is most effective. This explains why PSO and GA show a larger relative speedup in the five-condition setting than in the two-condition setting.

Third, even under these more favorable and more harmonized comparison settings, BOLT(32) still achieves the most favorable overall balance among accuracy, repeated-run stability, and computational cost. In particular, BOLT(32) reaches \(12.4 \pm 0.1\) mV, which is lower than PSO (2 cond.) (\(15.4 \pm 1.8\) mV), PSO (5 cond.) (\(14.1 \pm 1.2\) mV), GA (2 cond.) (\(14.5 \pm 3.3\) mV), and GA (5 cond.) (\(13.3 \pm 1.0\) mV). At the same time, its runtime remains only \(8.97 \pm 1.20\) s, far below even the JIT-accelerated PSO and GA baselines. It also requires only \(20636 \pm 3081\) model calls, which is substantially lower than all PSO and GA settings.

From a methodological perspective, this advantage can be attributed to the way BOLT allocates its computational budget. Instead of repeatedly evolving large populations across many iterations, BOLT performs a set of independent local refinements from diverse starting points and then applies multi-condition screening to select the most consistent candidate from the explored pool. This design reduces the number of required model evaluations while simultaneously improving the repeated-run stability of the final selected solution. Overall, the results show that the advantage of BOLT is not merely lower estimation error, but a more favorable joint trade-off among accuracy, computational efficiency, and repeated-run stability. Importantly, this advantage arises from the local-to-global calibration strategy of BOLT rather than from a single implementation-level acceleration technique, which makes the proposed framework particularly attractive for repeated electrochemical model calibration tasks requiring both speed and robustness.

The role of the main BOLT design choices can also be interpreted from the preceding comparisons. The candidate-number analysis in Section~4.2 shows that increasing \(N\) from 16 to 32 substantially improves repeated-run stability, while further increases provide only marginal gains. The two-condition and five-condition PSO/GA comparisons show that using broader condition coverage improves consistency but greatly increases model calls. BOLT exploits these observations by using only two informative profiles for local refinement and then using the remaining conditions for candidate screening. Thus, the observed performance is not caused by a single component, but by the combined effect of diversified initialization, fast local TRF refinement, batch-parallel execution, and multi-condition screening.

\subsection{Condition-wise Performance across Operating Conditions}

\begin{figure}[H]
\centering
\makebox[\textwidth][c]{
    \includegraphics[width=1.5\textwidth]{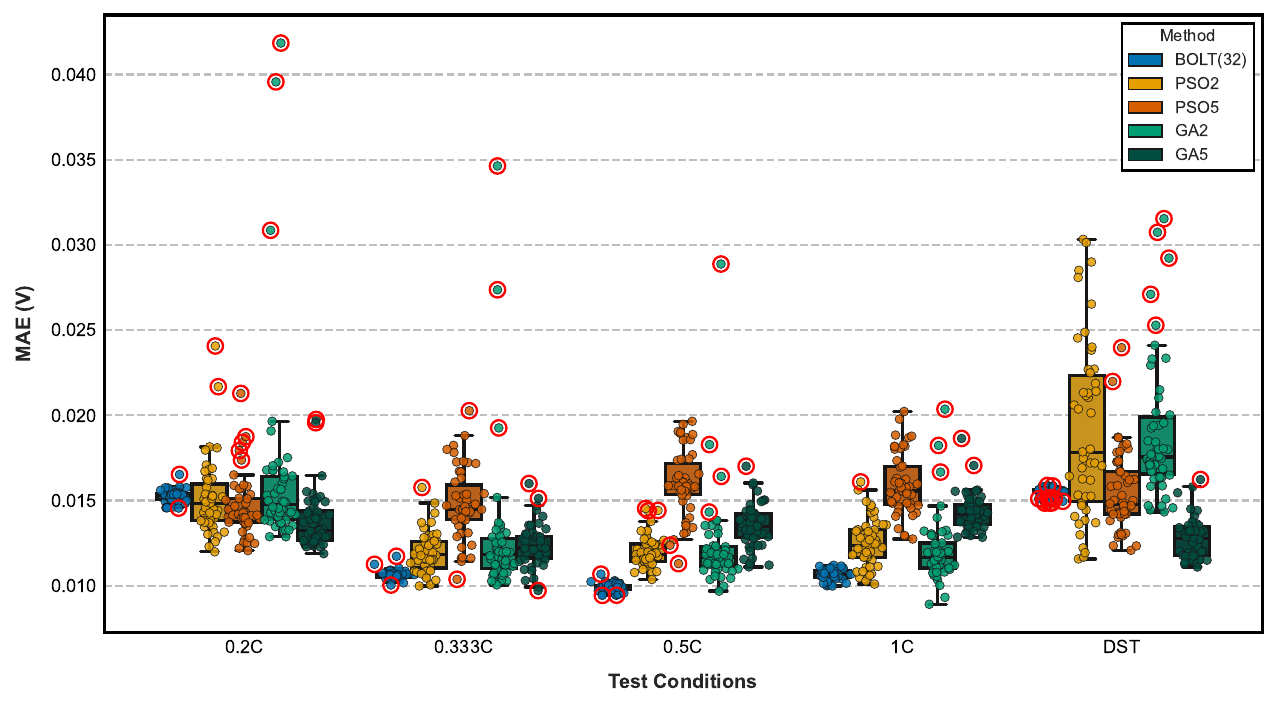}
}
\caption{Condition-wise MAE distributions of BOLT(32), PSO2, PSO5, GA2, and GA5 under the five operating conditions considered in this study: 0.2C, 0.333C, 0.5C, 1C, and DST.}
\label{fig:6}
\end{figure}

To further examine condition-wise behavior, Figure~\ref{fig:6} presents the
MAE distributions under the five operating conditions (0.2C, 0.333C, 0.5C,
1C, and DST) for BOLT(32), PSO (2 cond.), PSO (5 cond.), GA (2 cond.), and
GA (5 cond.). The figure explicitly shows the five operating conditions on
the horizontal axis and the five compared methods in the legend.

To complement the distributional comparison in Figure~\ref{fig:6},
Table~\ref{tab:conditionwise_voltage_mae} reports the numerical
condition-wise voltage MAE statistics under the five operating conditions.
This table provides a direct voltage-error comparison between the measured
terminal voltage and the model-predicted voltage under each condition, and
therefore makes the condition-wise fitting performance more explicit.

\begin{table}[H]
\centering
\small
\caption{Condition-wise voltage MAE statistics under different operating conditions. Values are reported as mean \(\pm\) standard deviation over 50 independent runs.}
\label{tab:conditionwise_voltage_mae}
\renewcommand{\arraystretch}{1.08}
\setlength{\tabcolsep}{3.8pt}
\begin{tabular}{@{}lccccc@{}}
\toprule
\textbf{Condition} &
\textbf{BOLT(32)} &
\textbf{PSO2} &
\textbf{PSO5} &
\textbf{GA2} &
\textbf{GA5} \\
\midrule
0.2C   & \(15.3 \pm 0.4\) & \(14.6 \pm 1.8\) & \(15.1 \pm 2.2\) & \(16.4 \pm 5.7\) & \(13.8 \pm 1.6\) \\
0.333C & \(10.6 \pm 0.3\) & \(14.8 \pm 2.0\) & \(12.0 \pm 1.2\) & \(12.7 \pm 4.1\) & \(12.4 \pm 1.2\) \\
0.5C   & \(9.9 \pm 0.2\)  & \(16.0 \pm 1.9\) & \(12.0 \pm 0.9\) & \(12.3 \pm 2.8\) & \(13.5 \pm 1.3\) \\
1C     & \(10.7 \pm 0.3\) & \(15.9 \pm 1.8\) & \(12.5 \pm 1.4\) & \(12.1 \pm 2.0\) & \(14.3 \pm 1.1\) \\
DST    & \(15.5 \pm 0.3\) & \(15.5 \pm 2.4\) & \(19.1 \pm 5.1\) & \(19.0 \pm 4.1\) & \(12.8 \pm 1.2\) \\
\midrule
Avg.   & \(12.4 \pm 0.1\) & \(15.4 \pm 1.8\) & \(14.1 \pm 1.2\) & \(14.5 \pm 3.3\) & \(13.3 \pm 1.0\) \\
\bottomrule
\end{tabular}

\vspace{2pt}
\begin{minipage}{0.96\linewidth}
\scriptsize
PSO2/GA2 denote the two-condition setting using 0.5C and 1C during optimization, whereas PSO5/GA5 denote the five-condition setting using 0.2C, 0.333C, 0.5C, 1C, and DST. The voltage MAE is computed between the measured terminal voltage and the model-predicted voltage under each operating condition. All values are reported in mV.
\end{minipage}
\end{table}

As shown in Table~\ref{tab:conditionwise_voltage_mae}, BOLT(32) achieves
the lowest average voltage MAE across the five operating conditions, with an
Avg. MAE of \(12.4 \pm 0.1\) mV. More importantly, BOLT(32) shows much
smaller standard deviations than the PSO and GA baselines, indicating stronger
repeated-run stability. At the condition level, BOLT(32) gives the best mean
MAE under 0.333C, 0.5C, and 1C, and remains competitive under 0.2C and DST.
Although GA5 achieves lower mean errors under 0.2C and DST, its average error
over all five conditions remains higher than that of BOLT(32). Similarly, PSO5
improves the overall average error compared with PSO2, mainly by reducing the
errors under 0.333C, 0.5C, and 1C, but it still exhibits a relatively large
error and variance under the DST profile.

These results are consistent with the box-plot distributions in
Figure~\ref{fig:6}. Compared with the population-based baselines, BOLT(32)
does not simply improve one specific operating condition; instead, it provides
a more balanced condition-wise performance with substantially stronger
run-to-run robustness. This confirms that the proposed multi-start local
refinement and multi-condition screening strategy can identify parameter
sets that maintain stable voltage-fitting accuracy across the available
operating-condition matrix.

The condition-wise results also clarify the role of condition coverage during
optimization. The five-condition PSO and GA settings generally improve the
average fitting accuracy compared with their two-condition counterparts, but
this improvement is obtained at the cost of substantially higher model calls
and runtime, as reported in Table~\ref{tab:3}. In contrast, BOLT(32) uses only
0.5C and 1C for local refinement, while the additional operating conditions
are used for candidate screening. Therefore, the results in Figure~\ref{fig:6}
and Table~\ref{tab:conditionwise_voltage_mae} should be interpreted as evidence
of stronger cross-condition consistency rather than strict out-of-sample
generalization. Under this interpretation, BOLT(32) provides a practical
advantage by identifying parameter sets that remain accurate and stable across
the available operating-condition matrix while requiring far lower
computational cost than the five-condition PSO and GA baselines.

Overall, Figure~\ref{fig:6} and Table~\ref{tab:conditionwise_voltage_mae}
lead to a consistent conclusion. The practical value of BOLT lies not only in
fast computation, but also in its ability to maintain stable voltage-fitting
quality across repeated runs and multiple operating conditions. This combination
of computational efficiency, repeated-run stability, and cross-condition
consistency is particularly important for electrochemical model calibration
tasks in which parameter estimation must be performed repeatedly and reliably.

\subsection{Computational complexity, scalability, and deployment implications}
\label{sec:complexity_deployment}

The practical runtime of BOLT is determined not only by the total number of electrochemical-model evaluations, but also by how the independent local-refinement tasks are scheduled across parallel workers. Therefore, the complexity discussion below focuses on the wall-clock cost under the batch-parallel execution structure used in BOLT, rather than only on the serial total computational work.

In the following analysis, \(M_{\mathrm{fit}}\) denotes the number of operating profiles in the fitting data set \(\mathcal{D}_{\mathrm{fit}}\), and \(M_{\mathrm{screen}}\) denotes the number of operating profiles in the screening data set \(\mathcal{D}_{\mathrm{screening}}=\{\mathcal{C}_{m}\}_{m=1}^{M_{\mathrm{screen}}}\). Let \(N\) be the total number of initial candidate vectors, \(n\) the number of available virtual CPUs, and \(J=\lceil N/n\rceil\) the number of parallel scheduling batches. In the present study, \(N\) is selected as an integer multiple of \(n=16\), so \(J=N/n\), and each batch contains exactly \(n\) candidates. The \(j\)-th batch is denoted by \(\mathcal{G}_{j}\). Let \(n_{\theta}\) be the dimension of the estimated parameter vector, \(L_i\) the number of TRF iterations required by the \(i\)-th candidate, \(\gamma_{\mathrm{TRF}}(n_{\theta})\) the average model-evaluation factor per TRF iteration, and \(\tau_{\mathrm{sim}}(T)\) the computational cost of one model simulation over a profile with \(T\) time steps. For finite-difference sensitivity calculations, \(\gamma_{\mathrm{TRF}}(n_{\theta})\) generally increases with \(n_{\theta}\), because additional model evaluations are required to approximate the parameter sensitivities.

The local-refinement cost of the \(i\)-th candidate can be approximated as
\begin{equation}
\mathcal{T}_{i}^{\mathrm{fit}}
=
\mathcal{O}\left[
L_i \gamma_{\mathrm{TRF}}(n_{\theta})
\sum_{m=1}^{M_{\mathrm{fit}}}
\tau_{\mathrm{sim}}\!\left(T_{m}^{\mathrm{fit}}\right)
\right],
\label{eq:candidate_local_cost}
\end{equation}
where \(T_{m}^{\mathrm{fit}}\) is the number of time steps in the \(m\)-th fitting profile. This expression accounts for the repeated model simulations required during TRF-based local refinement on \(\mathcal{D}_{\mathrm{fit}}\).

Because the candidates within each batch are refined in parallel, the wall-clock cost of one batch is governed by the slowest candidate in that batch rather than by the sum of all candidate costs. Therefore, the batch-parallel local-refinement wall-clock cost can be expressed as
\begin{equation}
\mathcal{T}_{\mathrm{local}}
=
\mathcal{O}\left[
\sum_{j=1}^{J}
\max_{i\in\mathcal{G}_{j}}
\left(
L_i \gamma_{\mathrm{TRF}}(n_{\theta})
\sum_{m=1}^{M_{\mathrm{fit}}}
\tau_{\mathrm{sim}}\!\left(T_{m}^{\mathrm{fit}}\right)
\right)
\right].
\label{eq:parallel_local_cost}
\end{equation}
This equation explicitly reflects the parallel-computing advantage of BOLT. Under balanced convergence behavior across candidates, the local-refinement wall-clock cost scales approximately with the number of scheduling batches \(J=N/n\), rather than directly with the total number of candidates \(N\). This is consistent with the implementation in which the tested candidate numbers \(N=16,32,48,64,\) and \(80\) correspond to \(J=1,2,3,4,\) and \(5\) complete batches on the 16-vCPU platform.

After local refinement, all refined candidates are evaluated on the screening data set \(\mathcal{D}_{\mathrm{screening}}\). This step only requires forward model simulations and does not involve repeated TRF iterations. Consistent with Algorithm~\ref{alg:bolt_ltg}, the conservative wall-clock cost of the screening step can be written as
\begin{equation}
\mathcal{T}_{\mathrm{screen}}
=
\mathcal{O}\left[
N
\sum_{m=1}^{M_{\mathrm{screen}}}
\tau_{\mathrm{sim}}\!\left(T_{m}^{\mathrm{screen}}\right)
\right],
\label{eq:screen_cost}
\end{equation}
where \(T_{m}^{\mathrm{screen}}\) is the number of time steps in the \(m\)-th screening profile. Therefore, the overall wall-clock cost can be summarized as
\begin{equation}
\mathcal{T}_{\mathrm{BOLT}}
=
\mathcal{T}_{\mathrm{local}}
+
\mathcal{T}_{\mathrm{screen}}
+
\mathcal{T}_{\mathrm{overhead}},
\label{eq:bolt_wall_clock_cost}
\end{equation}
where \(\mathcal{T}_{\mathrm{overhead}}\) includes process scheduling, memory transfer, result collection, and other implementation-level overheads. Because the local TRF refinement requires repeated residual and sensitivity evaluations, it is usually the dominant computational component. The screening step is comparatively cheaper because it only evaluates the already refined candidates through forward simulations.

Because the model-evaluation cost may vary across operating profiles, the profile-wise summation form is retained in Eqs.~\eqref{eq:candidate_local_cost}--\eqref{eq:screen_cost}, rather than assuming an identical average simulation cost for all profiles. In practice, the actual runtime is also affected by solver convergence variability across candidates, process scheduling, memory overhead, profile-length differences, and the efficiency of the Numba-accelerated model-evaluation routine.

This analysis also explains the candidate-number settings used in Section~4.2. Since the platform provides \(n=16\) vCPUs, the tested values \(N=16,32,48,64,\) and \(80\) correspond to complete parallel batches. This avoids incomplete final batches and improves utilization of the available computing resources. The sensitivity results show that increasing \(N\) from 16 to 32 substantially improves repeated-run stability, whereas further increases provide only marginal additional accuracy improvement at increased runtime. Therefore, \(N=32\) was selected as a practical balance between search diversity and computational cost.

This complexity analysis further clarifies the scalability of BOLT. For higher-fidelity electrochemical models such as SPMe or P2D models, the same local-to-global idea can be applied in principle, but the absolute identification time will inevitably increase because both \(\tau_{\mathrm{sim}}(T)\) and the effective parameter dimension \(n_{\theta}\) are larger. Therefore, direct extension to parameter-intensive P2D models would likely require additional acceleration strategies, such as parameter grouping, sensitivity-based parameter reduction, reduced-order discretization, or surrogate-assisted pre-screening. The relative benefit of BOLT is expected to remain most pronounced when repeated model evaluations dominate the total computational cost and can be parallelized efficiently.

From a deployment perspective, BOLT is mainly intended for rapid offline, cloud-side, or edge-side calibration rather than continuous execution on low-cost BMS microcontrollers. Once the parameters are identified, the resulting grouped SPM can be used in embedded BMS algorithms. Running the complete BOLT workflow on an actual BMS chip would require additional implementation work, such as C/C++ translation, memory optimization, and possibly fixed-point or hardware-specific acceleration. For this reason, the present work should be viewed as a calibration framework supporting BMS parameter updating and control-oriented battery digital twins, rather than as a fully embedded on-chip optimization implementation.

\subsection{Limitations and implications}
\label{sec:limitations_implications}

Although the present results demonstrate the practical advantages of BOLT on the grouped SPM and the tested 18650 NMC cell, several limitations should be noted. First, BOLT does not provide a formal guarantee of global convergence or parameter uniqueness. Electrochemical model calibration is generally nonconvex, ill-conditioned, and potentially non-identifiable, because different parameter combinations may produce similar voltage responses. Therefore, BOLT should be regarded as a practical calibration framework for improving repeated-run robustness and computational efficiency, rather than as a replacement for theoretical identifiability analysis. The synthetic known-parameter tests in Section~\ref{sec:synthetic_recoverability} complement the measured-data comparison by examining parameter recoverability and noise robustness under controlled model-consistent conditions.

Second, the measured-data results should be interpreted as cross-condition consistency over the available operating-condition matrix, rather than strict out-of-sample generalization. In BOLT, additional operating conditions are used for candidate screening because electrochemical parameter estimation benefits from informative current--voltage data to reduce profile-specific overfitting and improve practical identifiability. Further validation on fully held-out profiles, different cells, ageing states, and chemistries is still needed.

Third, the present validation is limited to a grouped SPM and one commercial 18650 NMC cell. The grouped SPM formulation improves computational tractability for control-oriented calibration, but it does not imply complete recovery of all individual physical parameters in a full P2D model. Extension to SPMe or P2D models would require additional parameter grouping, sensitivity-based parameter selection, reduced-order discretization, or other acceleration strategies. BOLT also differs from conventional multi-start optimization by combining batch-parallel local refinement with multi-condition screening, and differs from surrogate-assisted approaches by avoiding surrogate training while still relying on repeated electrochemical-model simulations.

Finally, the practical efficiency of BOLT depends on the available computing resources and implementation. As discussed in Section~\ref{sec:complexity_deployment}, BOLT is mainly intended for rapid offline, cloud-side, or edge-side calibration rather than continuous execution on low-cost BMS microcontrollers. Despite these limitations, BOLT provides a fast, repeatable, and physically interpretable calibration strategy for applications such as BMS parameter updating, control-oriented battery digital twins, and second-life battery screening.

\section{Conclusion}\label{sec6}

This paper presents BOLT, a batch-optimized local-to-global calibration framework for rapid and robust parameter estimation of electrochemical battery models. By combining efficient local refinement, diversified candidate exploration, and multi-condition consistency screening within a unified workflow, BOLT addresses two major practical limitations of conventional approaches: high computational cost and insufficient repeated-run robustness.

Experimental results based on a commercial 18650 lithium-ion cell and the grouped SPM show that BOLT achieves a more favorable trade-off among estimation accuracy, computational efficiency, and repeated-run stability than the PSO and GA baselines. In particular, BOLT(32) achieves an Avg\_MAE of \(12.4 \pm 0.1\) mV while requiring only \(20636 \pm 3081\) battery model calls and \(8.97 \pm 1.20\) s per run. Even when PSO and GA are evaluated under more demanding settings and with JIT-accelerated model evaluation, BOLT still maintains lower error, tighter repeated-run distributions, and substantially shorter runtime.

The synthetic-data validation further confirms the parameter recoverability of the proposed workflow under controlled conditions. In the noiseless case, BOLT recovers the known parameter vector at numerical precision. Under 1--3 mV voltage-noise perturbations, the reconstructed clean-voltage error remains far below the injected noise level, and the mean parameter ARE remains below 0.6\%.

The results further show that the total candidate number $N$ plays an important role in the accuracy--efficiency trade-off of BOLT. In the present study, $N=32$ provides the most favorable balance, while larger values of $N$ offer only marginal additional accuracy improvement at increased computational cost. This suggests that the proposed framework can be scaled in a controlled manner when more complex electrochemical models or larger parameter spaces are considered.

Overall, BOLT provides a practical solution for fast, stable, and repeatable electrochemical model calibration. Its ability to deliver low-error parameter estimates within seconds makes it promising for applications requiring repeated and reliable calibration, including BMS parameter updating, control-oriented battery digital twins, and second-life battery screening. Future work will extend the proposed framework to higher-fidelity electrochemical models, broader operating scenarios, and online parameter updating for embedded battery management applications.

\medskip

\textbf{Acknowledgments} \par
This work was supported by the Research Foundation - Flanders (FWO) (grant numbers 1252326N).

\medskip

\textbf{Declaration of competing interest}\par
The authors declare that they have no known competing financial interests or personal relationships that could have appeared to
influence the work reported in this paper.

\medskip

\textbf{Data availability}\par
The electrochemical battery model implementation used in this study is available at \texttt{https://github.com/FrankSuperG/CPG-SPMT}.

\medskip

\textbf{Author contribution} \par
Feng Guo: Conceptualization, Methodology, Software, Formal analysis, Investigation, Visualization, Writing – original draft, Writing – review \& editing.
Luis D. Couto: Conceptualization,  Writing – review \& editing.
Keivan Haghverdi: Writing – review \& editing.
Khiem Trad: Writing – review \& editing.
Grietus Mulder: Writing – review \& editing.

 \bibliographystyle{elsarticle-num}
 \bibliography{cas-refs}

\end{document}